# Comparison of Optimised Geometric Deep Learning Architectures, over Varying Toxicological Assay Data Environments

Alexander D. Kalian [1], Lennart Otte [2], Jaewook Lee [2], Emilio Benfenati [3], Jean-Lou C.M. Dorne [4], Claire Potter [5], Olivia J. Osborne [5], Miao Guo [2,*], Christer Hogstrand [6,*]

[1] Department of Nutritional Sciences, King's College London, London, United Kingdom
[2] Department of Engineering, King's College London, London, United Kingdom
[3] Istituto di Ricerche Farmacologiche Mario Negri IRCCS, Milan, Italy
[4] European Food Safety Authority (EFSA), Parma, Italy
[5] Food Standards Agency, London, United Kingdom
[6] Department of Analytical, Environmental and Forensic Sciences, King's College London, London, United Kingdom

* Corresponding authors:
Miao Guo (miao.guo@kcl.ac.uk); Christer Hogstrand (christer.hogstrand@kcl.ac.uk)

## Abstract

Geometric deep learning is an emerging technique in Artificial Intelligence (AI) driven cheminformatics, however the unique implications of different Graph Neural Network (GNN) architectures are poorly explored, for this space. This study compared performances of Graph Convolutional Networks (GCNs), Graph Attention Networks (GATs) and Graph Isomorphism Networks (GINs), applied to 7 different toxicological assay datasets of varying data abundance and endpoint, to perform binary classification of assay activation. Following pre-processing of molecular graphs, enforcement of class-balance and stratification of all datasets across 5 folds, Bayesian optimisations were carried out, for each GNN applied to each assay dataset (resulting in 21 unique Bayesian optimisations). Optimised GNNs performed at Area Under the Curve (AUC) scores ranging from 0.728-0.849 (averaged across all folds), naturally varying between specific assays and GNNs. GINs were found to consistently outperform GCNs and GATs, for the top 5 of 7 most data-abundant toxicological assays. GATs however significantly outperformed over the remaining 2 most data-scarce assays. This indicates that GINs are a more optimal architecture for data-abundant environments, whereas GATs are a more optimal architecture for data-scarce environments. Subsequent analysis of the explored higher-dimensional hyperparameter spaces, as well as optimised hyperparameter states, found that GCNs and GATs reached measurably closer optimised states with each other, compared to GINs, further indicating the unique nature of GINs as a GNN algorithm.

## Keywords

# Graphical Abstract

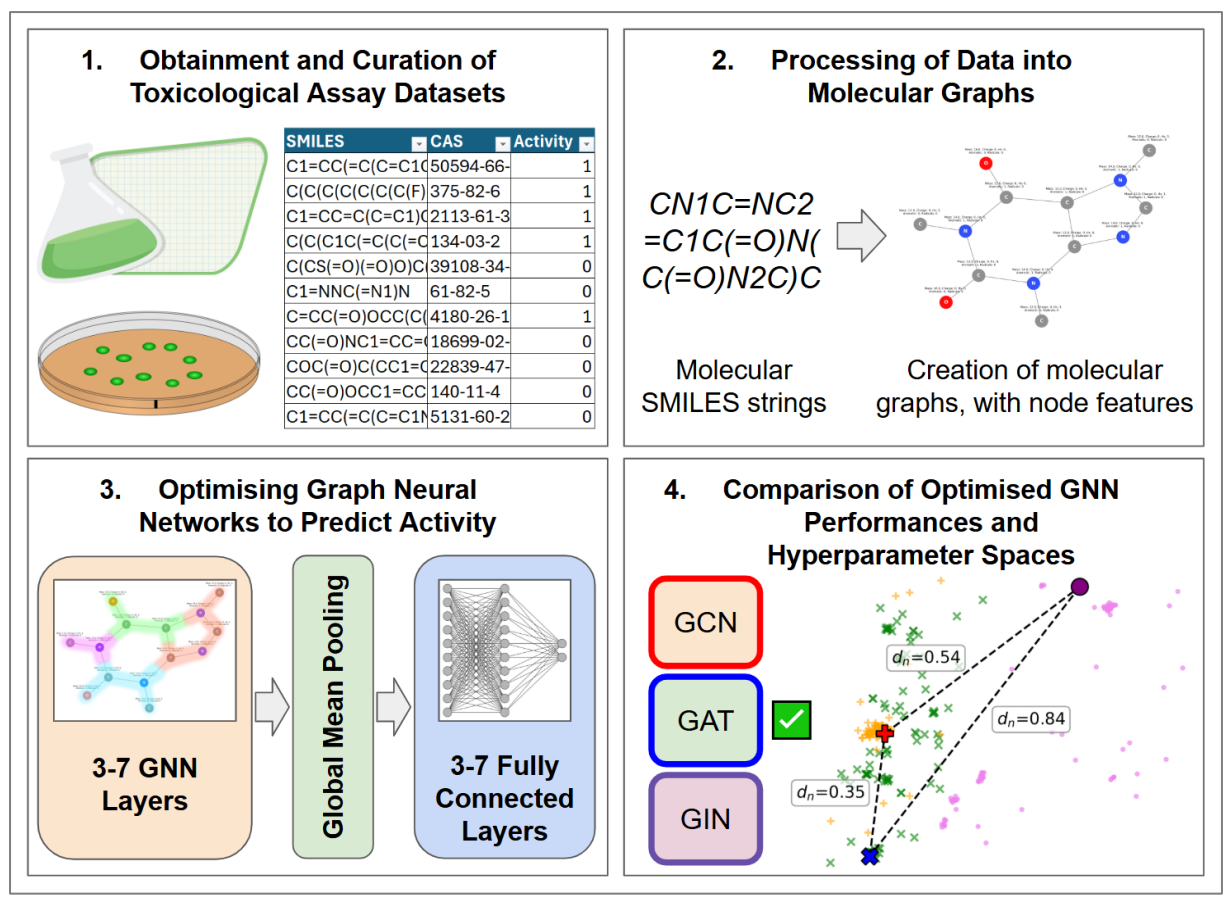

*Graphical Abstract: Visual overview of the research pipeline used in this study.*

# 1 – Introduction

## 1.1 – Geometric Deep Learning in the Toxicological Sciences

Geometric deep learning entails deep learning on non-Euclidean data structures, such as graphs and manifolds [1,2]. This differs from conventional deep learning algorithms, which primarily process grid-structured matrices of data (e.g. image pixel data, or matrices representing sequences of text) [1,2]. Geometric deep learning is emerging in application to data science in computational toxicology (as well as the wider field of cheminformatics), as approaches such as Graph Neural Networks (GNNs) may directly process molecules as molecular graphs which encode their native structures of bonded atoms, enriched with further physicochemical information about constituent atoms and bonds, via node and edge feature vectors [1-5]. GNNs hence enable seamless training, testing and predicting across molecular datasets, with algorithms that conserve the inherently graph-structured form of molecules [1-5]. GNNs have emerged as an especially effective technique in Quantitative Structure-Activity Relationship (QSAR) modelling [3-5], which may be used to improve human, animal and environmental health, via safer and more effective chemicals, while simultaneously helping reduce the need for animal testing [4].

Beyond molecular graphs, GNNs are also prominent in use for predicting over knowledge graphs -general relational data structures that encode entities and their relations [1,2]. This is also used in cheminformatics studies [6,19], however molecular graph-based approaches remain more typical and intuitively suited for chemical data representation [1-5].

A variety of different GNN architectures have been invented, by foundational researchers in Artificial Intelligence (AI), typically designed for broad applicability across different domains [7,8,10]. A common theme across GNN variants is the paradigm of message-passing, in which node (and potentially edge) features are altered to contain information from neighbours, to encode information about the surrounding connected environment and hence empower a model to de-facto discern the wider graph topology [1,2,7,8,10].

### 1.2 – Graph Convolutional Networks

Graph Convolutional Networks (GCNs) were introduced in 2016 by T.N. Kipf and M. Welling [7], implementing a form of message-passing analogous to convolutions in Convolutional Neural Networks (CNNs – which are widely used for processing Euclidean-structured image data). These graph convolutions use matrix multiplication, to pass node features between immediate neighbours, but may be carried out *k* times, through *k* subsequent GCN layers, to spread information across *k*-hop node neighbours in a graph [7]. A trainable weights matrix is simultaneously used, to optimise the influence of different specific node features, for empowering the most effective predictions [7].

A schematic diagram of GCN message passing may be found below, in Fig. 1:

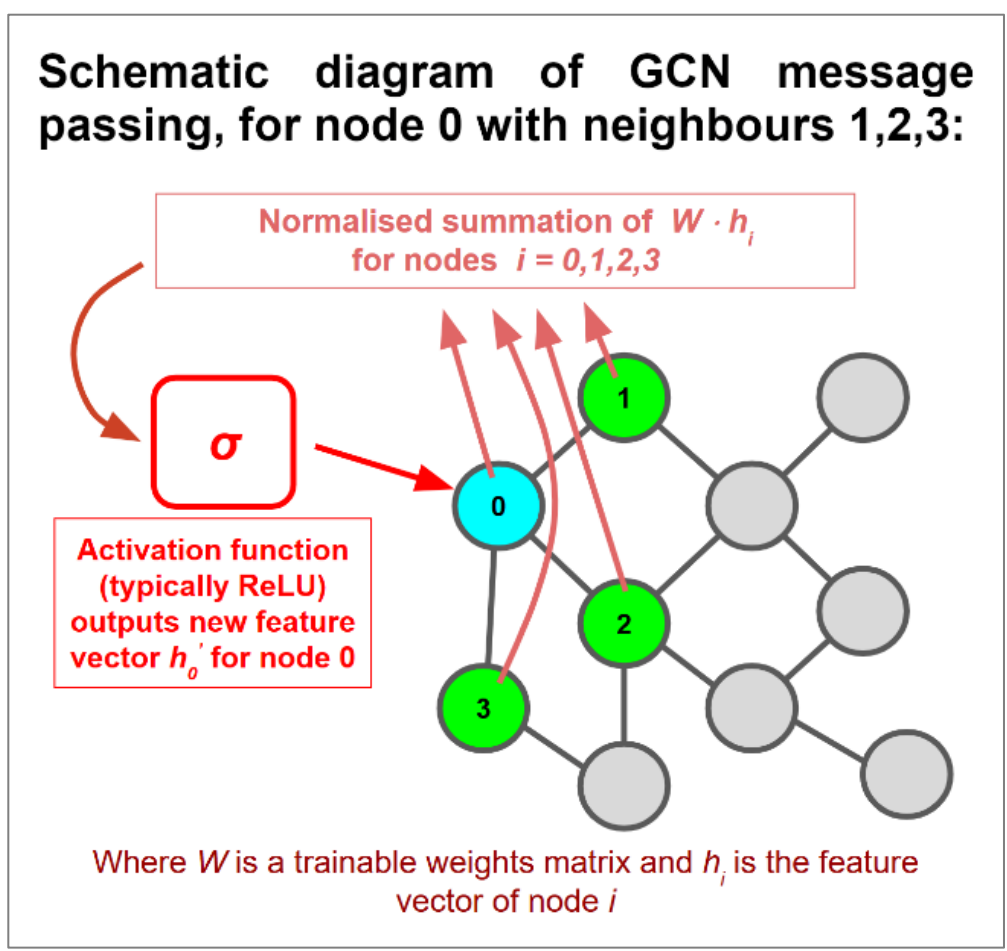

*Figure 1: Schematic overview of GCN message passing, for a given node in a graph.*

The GCN message passing algorithm, as shown in Fig. 1, may be formally defined via the following propagation rule between GCN layers [7]:

$$H^{(l+1)} = \sigma\left(\widetilde{D}^{-\frac{1}{2}}\widetilde{A}\widetilde{D}^{-\frac{1}{2}}H^{(l)}W^{(l)}\right) \quad (1)$$

Where $H^{(l)}$ is the node feature matrix for a given graph, at the $l^{th}$ GCN layer (starting from an unmodified node feature matrix at $l = 0$). $\sigma$ is an activation function - typically Rectified Linear Unit (ReLU). $\widetilde{D}^{-\frac{1}{2}}\widetilde{A}\widetilde{D}^{-\frac{1}{2}}$ represents a normalised adjacency matrix for that graph, with self-looping for individual nodes, where $\widetilde{D}_{ii} = \sum_j \widetilde{A}_{ij}$ (a diagonal degree matrix for the graph) and $\widetilde{A} = A + I_N$ (the graph's adjacency matrix $A$ added to its identity matrix $I_N$, to enable self-looping). $W^{(l)}$ is a weights matrix for layer $l$, which is optimised via gradient descent.

**1.3 – Graph Attention Networks**

Graph Attention Networks (GATs) were introduced in a 2017 arXiv publication (formally published in ICLR in 2018) by P. Veličković et al. [8]. GATs were designed to improve on the graph convolution based message-passing algorithm of GCNs, by instead leveraging self-attention [8] – used successfully in defining the transformer model architecture for Natural Language Processing (NLP) applications, in the 2017 publication "Attention Is All You Need" by A. Vaswani et al. [9].

A schematic diagram of GAT message passing may be found below, in Fig. 2:

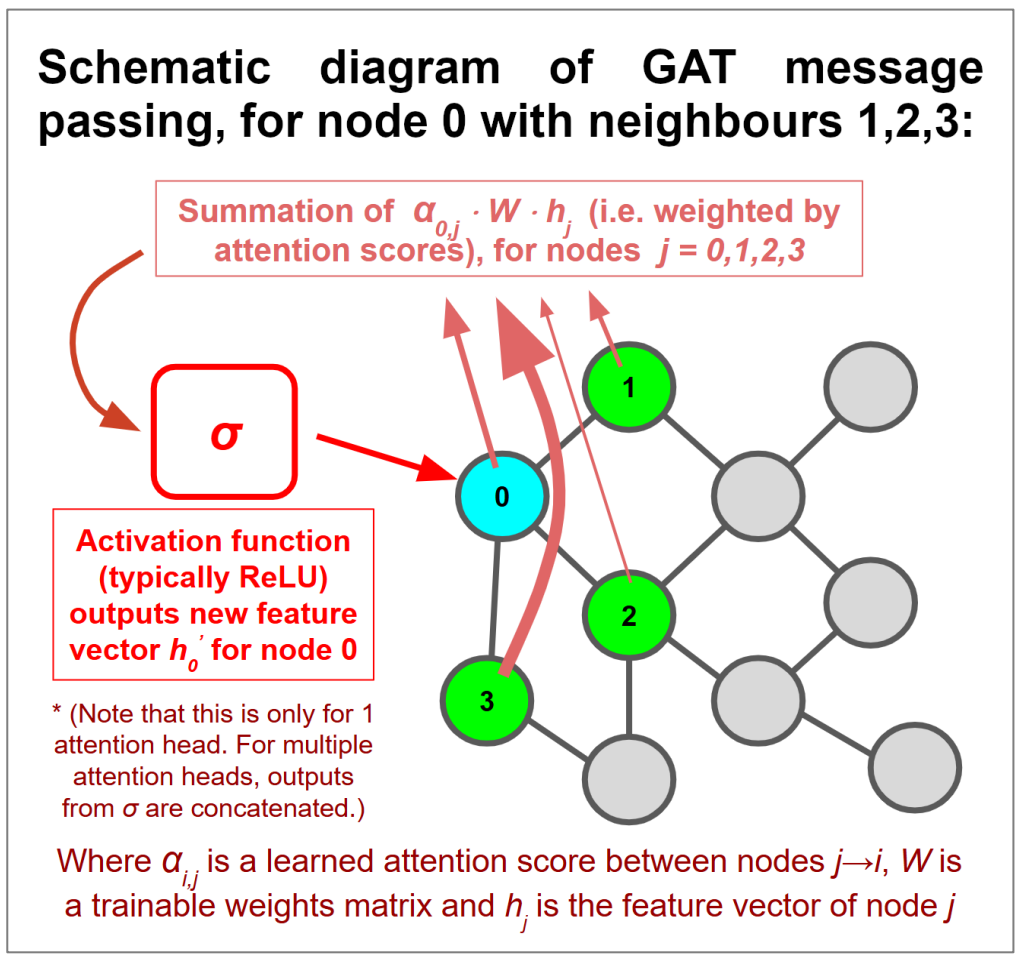

*Figure 2: Schematic overview of GAT message passing, for a given node in a graph.*

As per Fig. 2, the use of learned attention scores is a key difference between GAT message passing and classic GCN message passing.

Self-attention may be understood as a mechanism for capturing context-dependent relationships within sets or sequences of data, which can span over longer ranges between elements that are non-adjacent, rather than simply assuming that adjacent elements are more closely related [8,9]. This is achieved by using matrices of learned attention weights, which capture context-specific dependencies and relationships across different elements of a sequence or set (e.g. context-dependent relationships between different words of an input sentence, for NLP applications) [8,9].

In the context of GATs, self-attention mechanisms are used for message passing, hence enabling the model to learn more complex representations of how different nodes of a graph are related to each other [8]. For computational toxicology applications, GATs may hence pose a more sophisticated architecture than GCNs, for probabilistic modelling how different atomic nodes within molecular graphs collectively influence resulting toxicological properties [4,6] (e.g. forming different substructures, giving rise to hydrophobic/lipophilic properties, potentially amplifying each other's effects, potentially cancelling out each other's effects etc.).

Further to this, multiple attention heads may be used – i.e. self-attention is performed multiple times in parallel, using independently initialised attention weights matrices, enabling multiple parallel convergences to different beneficial optima in the trained attention weights space [8]. This allows the GAT to potentially capture and exploit a diversity of uniquely useful context-dependent relationships in the graph data [8]. Updated node features differ between multiple attention heads, hence the outputs are typically averaged [8].

The GAT message passing algorithm, of Fig. 2, may be formally defined by the following layer-wise propagation rule [8]:

$$h_i^{(l+1)} = ||_{m=1}^{M} \sigma \left( \sum_{j \in \mathcal{N}(i)} \alpha_{ij}^{(l,m)} W^{(l,m)} h_j^{(l)} \right) \quad (2)$$

Where $h_i^{(l+1)}$ is a new node feature vector for node $i$ at layer $(l+1)$, $M$ is the number of attention heads, $\sigma$ is a non-linear activation function - typically Exponential Linear Unit (ELU) or ReLU, $\alpha_{ij}^{(l,m)}$ is the trainable attention weight relating node $i$ to node $j$ (for attention head $m$ and at layer $l$), $W^{(l,m)}$ is a trainable weights matrix (for attention head $m$ and at layer $l$) which linearly transforms the feature vector into a new embedding and $h_j^{(l)}$ is the node feature vector for node $j$ at layer $(l)$. Please note that $||_{m=1}^{M}$ denotes concatenation across all attention heads. For the final GAT layer, outputs from multiple attention heads are however instead averaged.

Attention weights are configured via [8]:

$$a_{ij}^{(l,m)} = softmax_i \left( e_{ij}^{(l,m)} \right) \quad (3)$$

*where:*

$$e_{ij}^{(l,m)} = LeakyReLU \left( a^{(l,m)^T} \left[ W^{(l,m)} h_i^{(l)} || W^{(l,m)} h_j^{(l)} \right] \right) \quad (4)$$

Where $e_{ij}^{(l,m)}$ is an unnormalized attention weight (normalised into $a_{ij}^{(l,m)}$ via a softmax function) and $a^{(l,m)^T}$ is the transpose of a vector $a^{(l,m)}$ containing trainable parameters which control how the node features of nodes $i$ and $j$ are combined.

Similarly to GCNs, as described in Section 1.2, GAT message passing exclusively acts over immediate neighbours, and requires multiple *k* iterations, to result in message passing over *k*-hop neighbouring nodes [8].

## 1.4 – Graph Isomorphism Networks

Graph Isomorphism Networks (GINs) were introduced in a 2019 publication by K. Xu et al. [10]. GINs are designed to improve GNN expressiveness and ability to distinguish between subtly different graphs, particularly in the case of graph isomorphism, where a variety of conventional GNNs have been demonstrated to fail to distinguish between certain non-isomorphic graphs [10]. The introduced GIN algorithm has been demonstrated to be equally as powerful as the 1-Weisfeiler-Lehman (1-WL) test for graph isomorphism [10].

This is especially important in computational toxicology and wider cheminformatics applications, using GNNs to process molecular graphs, given that seemingly minor differences in molecular substructures are known to sometimes influence vastly different biological effects.

GINs navigate message passing, for a given node, by summing a scaled version of the node's features with the features of its immediate neighbours, before then applying a Multi-Layer Perceptron (MLP – i.e. a conventionally used feed-forward neural network) to process the summed result into a new node feature vector [10]. In this study, the MLP used by the GIN architecture will be referred to as a $MLP_{GIN}$.

A schematic diagram of GIN message passing may be found below, in Fig. 3:

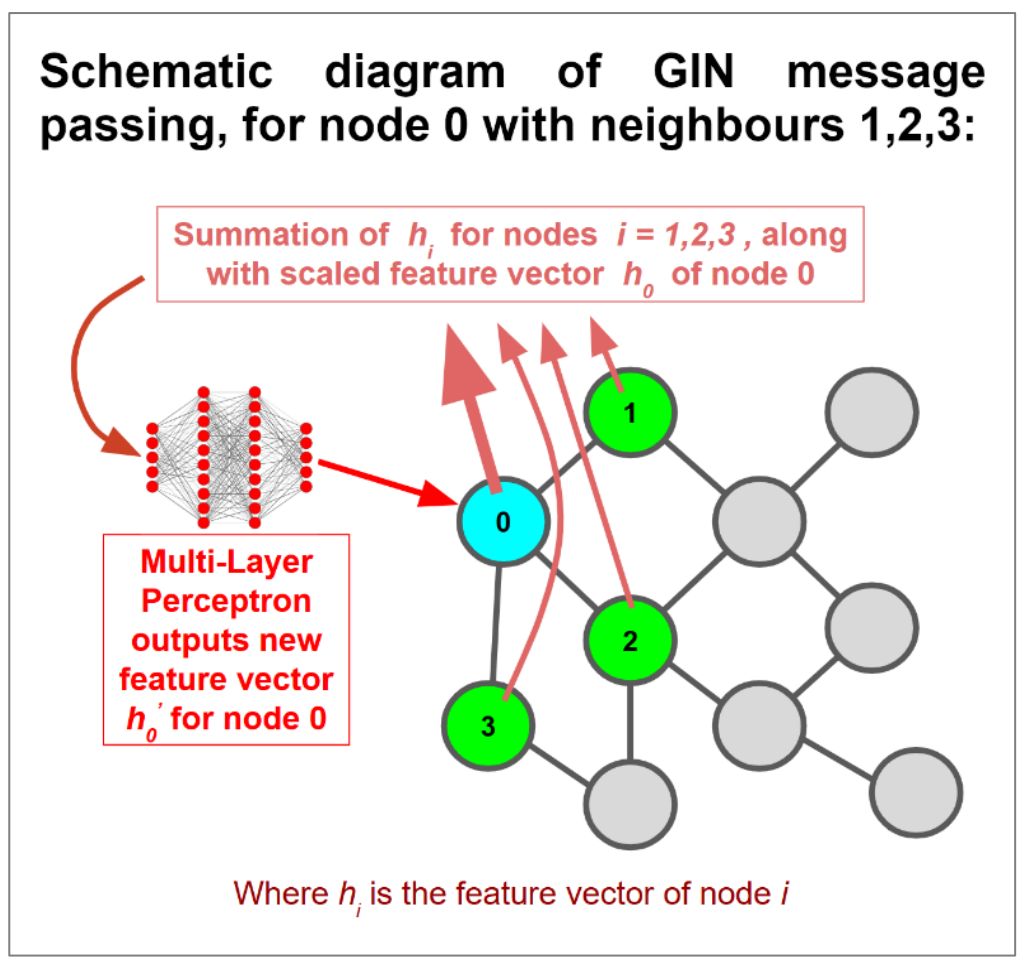

*Figure 3: Schematic overview of GIN message passing, for a given node in a graph.*

GIN message passing, as per Fig. 3, may be formally defined by the following layer-wise propagation rule [10]:

$$h_i^{(l+1)} = MLP_{GIN}^{(l)}\left((1 + \epsilon^{(l)})h_i^{(l)} + \sum_{j \in \mathcal{N}(i)} h_j^{(l)}\right) \quad (5)$$

Where $MLP_{GIN}^{(l)}$ represents a MLP$_{GIN}$ at layer $l$, $\epsilon^{(l)}$ represents a trainable parameter which governs the scaling of node $i$'s features $h_i^{(l)}$ at layer $l$ and $h_j^{(l)}$ represents the features of a neighbouring node $j$.

Note that the MLP$_{GIN}$ contains its own trainable parameters, and may be defined via the following layer-wise propagation rule [11]:

$$x^{(L+1)} = \sigma\left(W^{(L+1)}x^{(L)} + b^{(L)}\right) \quad (6)$$

Where $x^{(L+1)}$ is the feature vector representation at layer $(L + 1)$, $\sigma$ is an activation function, $W^{(L+1)}$ is a trainable weights matrix for layer $(L + 1)$, $x^{(L)}$ is the previous feature vector representation at layer $L$ and $b^{(L)}$ is a trainable bias vector for layer $L$.

Similarly to GCNs, as described in Section 1.2, GIN message passing exclusively acts over immediate neighbours, and requires multiple *k* iterations, to result in message passing over *k*-hop neighbouring nodes [10].

### 1.5 – Overview of GNNs Used in Computational Toxicology Studies

GCNs, GATs, GINs and other GNN architectures have found application in various computational toxicology studies [1-6,12-31].

GCNs have been widely used to classify mutagenicity (a toxicological endpoint regarding whether or not molecules cause genetic mutations) in various studies, such as a 2021 study by H. Gini et al. which also obtained relevant molecular substructures [12], a 2022 study by P. Fradkin et al. which predicted both mutagenicity and carcinogenicity – leveraging both to output final carcinogenicity predictions [13], as well as a 2023 study by H.J. Moon et al. which used subgraph embeddings (rather than whole molecular graphs) for predicting for aromatic hydrocarbons [14].

Furthermore, GCNs have been applied to other endpoints, such as in a 2021 study by J. Romano et al. which used GCNs to predict over various toxicological assays [15], a 2019 study by C. Cai et al. which used numerous methods (including GCNs) to predict drug-induced cardiotoxicity [16], as well as a 2024 study by L. Zhang et al. which used GCN layers (as a constituent block of a larger AI model) to predict molecular initiating events on protein targets relevant to neurotoxicity [17], among other studies.

GATs have similarly been applied to computational toxicological research. A 2023 study by H. Wang et al. used GATs to predict over 15 different endpoints, including ecotoxicology endpoints such as persistence in water and sediment, as well as biotransformation half-life in fish [18]. Our previous research (a 2024 study) used GATs to predict neurotoxicity, developmental toxicity and reproductive toxicity of brominated flame retardants [19]. In addition to this, a 2023 study by Y. Tong et al. used GATs to predict small molecule toxicity, over a variety of human target proteins [20].

GINs have experienced broad application over a range of toxicological studies. A 2023 study by G. Wang et al. used GINs to predict over a variety of toxicological endpoints: hepatotoxicity, cardiotoxicity, irritation and corrosion, respiratory toxicity and endocrine disruption [21]. Conversely, a 2020 study by Y. Peng et al. used GINs to predict absorption, distribution, metabolism and excretion-toxicity properties of molecules [22]. Furthermore, a 2024 study by

Q. Yu et al. used GINs to classify peptide toxicity (in the context of general toxicity to animals, e.g. toxins found in spider venom) [23].

Computational toxicological studies have also utilised other GNN architectures, such as a 2024 study by J.N. Ren et al. which used Graph Transformer Networks (GTNs, which leverage attention mechanisms similarly to GATs, but with attention applied over the whole graph for message passing for a given node, not just for immediate neighbours) to predict reproductive toxicity [24], while a 2023 study by J. Liu et al. used GraphSAGE (a type of GNN), alongside a bidirectional recurrent gated neural network (used as a language model), to predict over various endpoints from the Tox21 dataset [25].

### 1.6 – Comparison Studies of GNN Architectures in Computational Toxicology

Despite the general advantages and disadvantages of specific GNN architectures being known, as per foundational AI research studies [7,8,10], the more nuanced implications for use in computational toxicology and wider cheminformatics applications are poorly explored. Although a significant number of peer-reviewed GNN-based studies exist, in the computational toxicology and wider cheminformatics research spaces, with results readily available, the specific endpoints, datasets, other AI model blocks, validation methods, optimisation methods and other considerations differ [26-31], hence enabling direct comparisons between the inherent GNN architectures invalid.

There have been a limited number of studies that have attempted to draw more direct comparisons between different GNN architectures, in this field of research. A 2023 study, by R. Ketkar et al., compared GCNs, GATs, GTNs and Attentive FP (another GNN architecture) in their performances over toxicity data related to fish, Daphnia magna, Tetrahymena pyriformis, and Vibrio fischeri [26]. Although the study found Attentive FP to be the highest performing architecture, the authors acknowledged that future research would benefit from incorporating Bayesian optimisation of hyperparameters between the GNNs, to lead to more valid comparisons between GNNs in their optimised states [26].

Similarly, a 2025 study by S. Monhem et al., developed a novel GNN architecture and compared it to GCNs, GATs, GINs (including enhanced GINs) and Graph Total Variation (GTV), for training and testing on 8 toxicity datasets: the Tox21 dataset, Ames mutagenicity, skin sensitisation, carcinogenicity, drug-induced liver injury, acute toxicity median lethal dose and drug-mediated blockade of the voltage-gated potassium channel [27]. While the 2025 study found the novel GNN to outperform the other architectures, no hyperparameter optimisation technique was performed between the different architectures, to enable a comparison between optimised models [27].

A 2024 review article by M. Besharatifard and F. Vafaee, compared GNN-based models from studies which predicted synergistic drug combinations [28]. The GNNs considered included GCNs, Graph Autoencoders (GAEs), GATs and Graph Sample and Aggregate (GraphSAGE) [28]. Although models from 25 different studies were compared, across various performance metrics, underlying methods and more, the ability to infer the suitability of the underlying GNN architectures themselves, from direct comparisons between these differently configured, trained, tested and optimised models, is limited [28].

Other review articles, evaluating GNN-based models from a range of studies (which inherently used differing datasets, model configurations, validation techniques and more, hence limiting the ability to directly compare underlying GNN architectures), include a 2022 review by L.A. Alves et al., which directly addressed the emerging application of GNNs in virtual screening, exploring studies which used GNN architectures such as GCNs, GATs, GTNs, various forms of Message Passing Neural Networks (MPNNs) and more [29]. Additionally, a 2022 review by

Z. Zhang et al. explored the research space for GNNs applied to predicting drug-target interactions, considering architectures such as GCNs, GATs, GINs, GraphSAGE and others [30]. Outside of the field of computational toxicology, yet still in the wider cheminformatics space, a 2022 review by P. Reiser et al. explored a wide range of GNN-based studies, across the domains of materials science and chemistry [31].

These studies demonstrate a growing application of GNNs to computational toxicology research, as well as the wider cheminformatics space [1-6,12-31], alongside an increasing research demand to compare the domain-specific implications of particular GNN architectures [26-31]. However, it is apparent from the current research landscape, that existing comparison studies have so far lacked the ability to draw direct valid conclusions between underlying GNN architectures, given largely arbitrary differences in model configuration, endpoint, dataset, training and testing considerations, validation methods and hyperparameter optimisation [26-31].

This study hence aims to address this research gap, by training, testing, hyperparameter optimising (via Bayesian optimisation) and comparing performance of GCNs, GATs and GINs, across a controlled range of different toxicological assay datasets, of varying data abundance.

## 2 – Materials and Methods

### 2.1 – Hypothesis Formulation

It is hypothesised that GCNs, GATs and GINs will be demonstrated by this study as effective GNN architectures for training and testing QSAR models of toxicological assay activation, in line with related studies, but that specific performances will differ between the architectures. GATs and GINs are hypothesised to overall outperform GCNs, given that GATs and GINs are designed as more advanced architectures, improving on older foundational GNN architectures such as GCNs. It is also hypothesised that specific GNN architectures will measurably outperform/underperform over environments of higher/lower data-abundance, as suitability for learning over abundant data, or alternatively suitability for data-carce learning applications, may naturally differ.

The optimised states of different GNN architectures are also hypothesised to naturally differ, even over shared or otherwise comparable types of hyperparameters, as the higher dimensional hyperparameter optimisation landscape is likely to be vastly complex, with numerous local maxima and minima. This realistically means that optimised models, reached through Bayesian optimisations of hyperparameters, are likely to at best be locally optimal and reached through search paths that were highly sensitive to initial conditions (e.g. randomly initialised trainable parameters, data used for training and testing etc.) which will naturally differ between the different GNN architectures (even when using a controlled random seed, as the architectures inherently differ in configuration of trainable parameters).

### 2.2 – Data Collection and Pre-Processing

7 toxicological assay datasets were selected from the CompTox Chemicals Dashboard [32], via stratified random sampling of datasets with a suitable class balance (for binary activity/inactivity of molecules - such that 45%-55% of molecules were active). The random sampling was purposefully stratified over a range of different dataset sizes, to gain a broad representation of differing data abundance/scarcity environments. Following selection, all assay datasets were curated via automated exclusion of inconsistent duplicate molecular datapoints, as well as automated conversion of CAS registry numbers into SMILES (Simplified

Molecular Input Line Entry System) strings, via the online database PubChem [33] (and exclusion of any datapoints with erroneous CAS registry numbers). Further to this, datasets were split into 5 folds of equal (or otherwise approximately equal) sizes, using stratified random sampling that ensured class balance within each fold.

The final selected datasets were (in ascending order of curated data abundance): NVS_ENZ_hBACE (241 molecules), CLD_CYP1A1_48hr (306 molecules), CEETOX_H295R_OHPROG (553 molecules), CCTE_Simmons_CellTiterGLO_HEK293T (560 molecules), LTEA_HepaRG_CYP2C19 (1012 molecules), LTEA_HepaRG_UGT1A1 (1013 molecules) and ATG_PXRE_CIS (3703 molecules).

Molecules were subsequently processed into molecular graphs, encoding atoms as nodes and bonds as edges, with atom-specific physicochemical properties (atomic number, atomic mass, electronegativity, dispersion coefficient, dipole polarizability, fusion heat, proton affinity and number of implicitly bonded hydrogen atoms) included as node features. Molecular graphs were ultimately stored via the PyG (PyTorch Geometric) [34] native data structure, for efficiently handling graph-structured data for geometric deep learning tasks.

## 2.3 – GNN Model Overview

All GNN-based QSAR models were implemented in Python, using the PyG and PyTorch [35] libraries. An overview of the generalised model architecture may be found in Fig. 4:

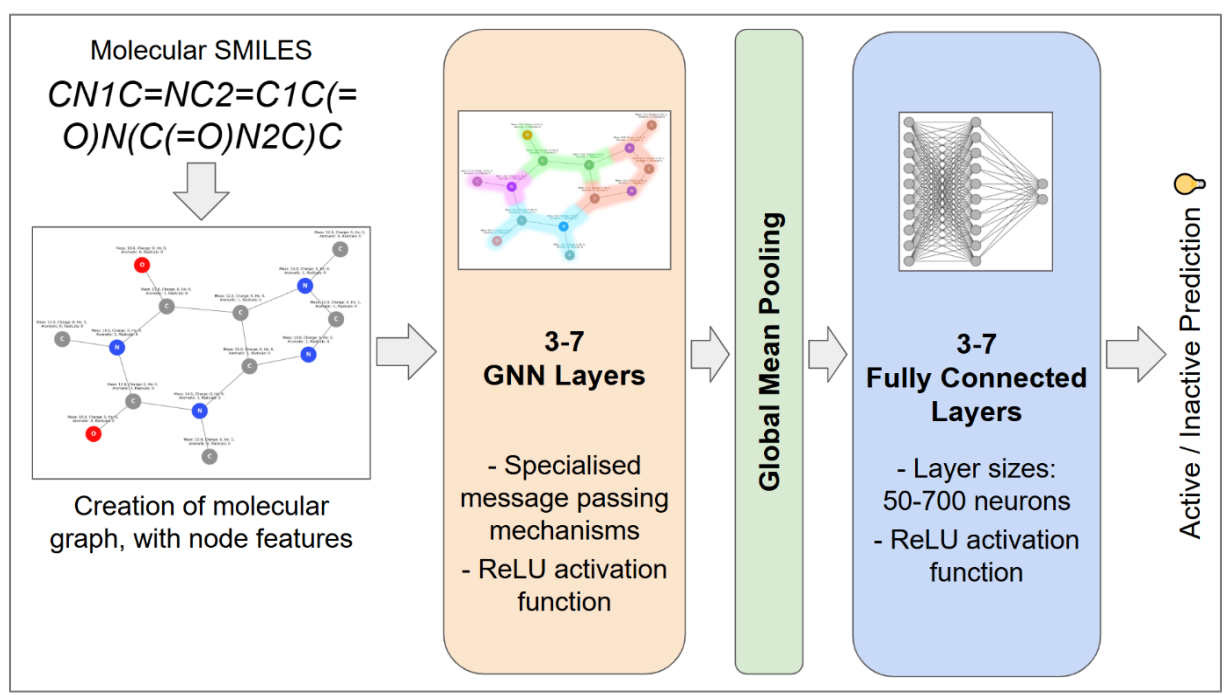

*Figure 4: Overview of the generalised model architecture used, for an example molecule.*

As per Fig. 4, each model first consisted of between 3-7 GNN layers (either GCN layers, GAT layers or GIN layers), designed to process input molecular graphs and carry out message passing through mechanisms specific to the particular type of GNN layer, generating node vectors that incorporated information from the surrounding graph-structured environment. The node vectors would subsequently be averaged into a single graph-level representation (i.e. a graph embedding), via a global mean pooling layer. Following this, 3-7 fully connected layers (de-facto MLP layers) were used to process graph embeddings, including a final layer of 2 neurons to output predictions of activity or inactivity (with 1 output logit for each binary class).

All GNN and fully connected layers used the ReLU activation function [36], while cross-entropy loss was used as the loss function, along with the Adam optimisation algorithm for gradient descent [37].

Model weights were randomly initialised and then trained for 500 epochs, repeated over 5-fold cross-validation, with model performance measured via both overall accuracy and ROC AUC (Receiver Operating Characteristic - Area Under the Curve) score. To compute ROC AUC score, the two output logits from the final layer were converted into class probabilities using the softmax function. The predicted probabilities for the positive class (i.e. active) were then compared with the ground truth labels, for ROC AUC score calculations.

Further to this, 5-fold cross validation was used – in each case, using a different fold as testing data.

**2.4 – Bayesian Optimisation**

The following table summarises the different hyperparameters, along with their search ranges, which were optimised via Bayesian optimisation:

| Hyperparameter Type | Min. Value | Max. value | Applies to GCN? | Applies to GAT? | Applies to GIN? |
|---|---|---|---|---|---|
| Number of GNN layers | 3 | 7 | ✅ | ✅ | ✅ |
| Number of hidden channels | 20 | 200 | ✅ | ✅ | ❌ |
| Number of attention heads | 2 | 10 | ❌ | ✅ | ❌ |
| Number of $MLP_{GIN}$ layers | 2 | 5 | ❌ | ❌ | ✅ |
| Size of $MLP_{GIN}$ layers | 50 | 500 | ❌ | ❌ | ✅ |
| Number of fully connected layers | 3 | 7 | ✅ | ✅ | ✅ |
| Size of fully connected hidden layers | 50 | 700 | ✅ | ✅ | ✅ |
| Dropout rate, for fully connected layers | 0.1 | 0.5 | ✅ | ✅ | ✅ |
| Learning rate | $1\times10^{-5}$ | $1\times10^{-3}$ | ✅ | ✅ | ✅ |
| Batch size | 8 | 256 | ✅ | ✅ | ✅ |

*Table 1: Summary of different explored hyperparameter types and search ranges, for the Bayesian optimisation.*

The chosen hyperparameters from Table 1 were selected to ensure flexibility for the number and size of layers shown in Fig. 2, while simultaneously considering other typically key hyperparameters such as learning rate, dropout rate and batch size. To enable a fair comparison between the different optimised GNNs, it was furthermore necessary to optimise unique hyperparameters for each GNN architecture which were key to the specific message passing involved; for GCNs (comparably the simplest of these GNN architectures), the number of hidden channels was sufficient, whereas for GATs, the number of attention heads (alongside the number of hidden channels) was deemed an appropriate hyperparameter to vary, while for GINs, the constituent aggregational MLP required optimisation both in terms of number and size of layers.

While the search ranges of these hyperparameters are to some extent arbitrary (as is inherent to a large number of Bayesian optimisation based studies [38, 39]), they were chosen to enable a sufficiently broad yet computationally feasible search range, while also remaining loosely in line with values that have been used in other models in our previous research, as well as other aforementioned studies outlined in the Introduction of this article.

The Bayesian optimisations were implemented via the Bayesian-optimization Python library [40], using 100 iterations, 10 initial points and a random state of 1 (with all other parameters at default settings). The Bayesian optimisations were configured to maximise ROC AUC score, averaged across all iterations of 5-fold cross-validation, for a given dataset. Bayesian optimisations were carried out for each given GNN architecture (out of 3 in total), training and testing over a given toxicological assay dataset (7 in total), hence 21 independent Bayesian optimisations took place in total.

**2.5 – Scope and Limitations of the Methodology**

This methodology differs from conventional approaches in QSAR modelling studies that carry out hyperparameter optimisations, in that training and testing sets were used (with optimisations configured to maximise ROC AUC score over the testing data), while devoid of any external validation set. External validation sets are typically used for the purpose of reducing bias and overfitting in the hyperparameter optimisation searches, to improve general applicability of the models to novel data (with an external validation set serving as unseen novel data for the models) [41, 42].

The scope of this study however is not in attempting to create QSAR models that are widely applicable to chemical space, but rather to explore the unique strengths and limitations of different GNNs in processing toxicological data of varying abundance. The Bayesian optimisations in this study hence serve a subtly different role to those in more conventional studies, in attempting to maximise GNN-based QSAR modelling performances over each toxicological dataset, to enable fair comparisons of the underlying strengths and weaknesses of different GNN architectures.

As comparisons of strengths and weaknesses of GNNs are the primary focus of this study, the dangers of overfitting to testing data are less of a concern than in conventional QSAR modelling studies. The risks of overfitting are however nonetheless mitigated, via use of 5-fold cross validation (as previously described), with Bayesian optimisations configured to maximise average ROC AUC score over the test sets across different fold-configurations. By closely integrating 5-fold cross validation into the performance metric being optimised, this methodology ensures that no single fold or partition of the data can disproportionately influence model selection, thereby reducing the risk of overfitting to any particular subset of the test data and supporting the reliability of performance trends observed across different GNN architectures.

While this does not eliminate the need for external validation in studies aiming for real-world QSAR model deployment, the methodology provides a robust internal framework for comparative architectural evaluation under consistent data conditions.

# 3 – Results and Discussion

**3.1 – Comparison of Optimised Model Performances**

A comparison table of peak AUC scores (averaged across all folds), for the different optimised GNN architectures, over the varying toxicological assay datasets, can be found below:

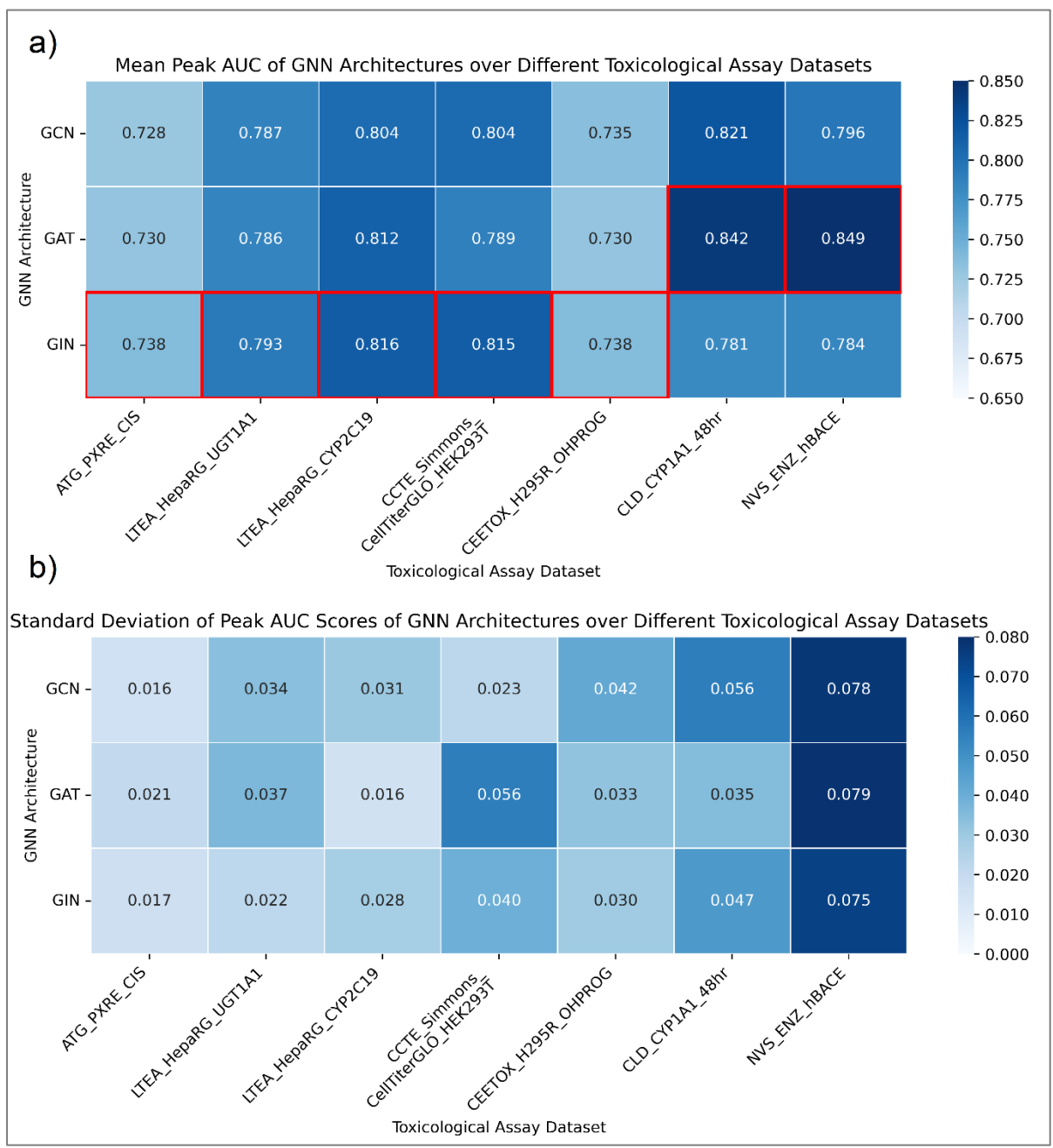

*Figure 5: Comparison table (with heat map colouring) of: a) mean peak AUC scores (averaged over all folds and given to 3 significant figures) of optimised GNN architectures, over differing toxicological data environments (assay datasets are ordered in descending data abundance, from left to right). For each dataset, the highest performing architecture is outlined in red. b) Standard deviation of the peak AUC scores used to calculate the mean peak AUC scores in (a).*

### 3.1.1 – Initial Findings

Fig. 5a displays a consistent trend where GINs outperform GCNs and GATs over the top 5 most data-abundant assays, whereas GATs outperform GCNs and GINs over the remaining top 2 most data-scarce assays.

The differences in mean peak AUC scores, between different GNNs trained and tested on the same dataset, are generally within close proximity. For examples, scores differ by less than 1 percentage point, for the ATG_PXRE_CIS, LTEA_HepaRG_UG1A1 and CEETOX_H295R_OHPROG datasets, and within 2 percentage points for the LTEA_HepaRG_CYP2C19 and CCTE_Simmons_CellTiterGLO_HEK293T datasets. This indicates that the trend of GIN advantages, in the high-data settings, may not be statistically robust. Conversely, the trend of GAT dominance in low-data environments is more pronounced (strongly dominant mean peak AUC scores).

It is however evident from Fig. 5b that the standard deviations, when taken as statistical uncertainties on the peak mean AUC scores of Fig. 5a, point to significant overlap in the model performance across architectures; this may hence call into question the statistical significance of any trends identified from Fig. 5a. Despite the high statistical uncertainties, the results in Fig. 5a outline GNN-specific trends that are consistent across all assays, hence a degree of significance may still be inferred. It should also be noted that Fig. 5b displays increased statistical noise in more data-scarce environments – this is to be expected, given that smaller datasets of molecules are more likely to be sensitive to specific partitions of training and testing data, hence leading to greater fluctuations in model performance, over different iterations of 5-fold cross validation.

The GCN architecture performed the weakest on average, out of all 3 considered GNN architectures. This may be attributed to GCNs comparably simpler design and less sophisticated message passing mechanisms in comparison with other two GNN architectures,. It is however less clear as to whether GINs or GATs are the most sophisticated of the considered architectures, based on theoretical foundations alone – and this is indeed reflected in the results of Fig. 5a, where GINs and GATs both displayed competing supremacy, over different data environments.

### 3.1.2 – Exploring the Differences Between GIN and GAT Relative Supremacy

#### 3.1.2.1 – Comparison of Expressiveness Between GINs and GATs

It may be suggested that GINs represented the most expressive GNN models used in this study, in terms of maximum number of trainable parameters, due to the use of MLPs within each GIN layer, where layer depths and widths are configurable via the hyperparameters in Table 1. Each $MLP_{GIN}$ held a maximum number of 5 layers, each with 500 neurons (analogous to hidden channels in GCNs and GATs – hence all 5 $MLP_{GIN}$ layers may be of size 500, when in secondary GIN layers or beyond). Given that an $MLP_{GIN}$ is fully-connected (and factoring in both weights and biases), this represents a maximum of 1,252,500 trainable parameters per $MLP_{GIN}$ – which may then be multiplied, when using additional GIN layers, as per Fig. 4. The GATs, in comparison, used a maximum of 9 attention heads per GAT layer, of which each held trainable parameters for processing up to 200 hidden channels. This hence represented a maximum sized weights matrix of 40,000 trainable weights, as well as a maximum size of 400 for the trainable attention vector, along with 200 bias terms – hence 40,600 trainable parameters per attention head, in total. For a maximum of 9 attention heads per GAT layer, this represents a combined total of 365,400 trainable parameters per layer, which is less than a third of single $MLP_{GIN}$.

Examination of the minimum configurations defined in Table 1 reveals significant contrasts. A minimal $MLP_{GIN}$ configuration (2 layers of 50 neurons each), contains approximately 5,100 parameters (factoring in both weights and biases), whereas a GAT layer with 2 attention heads and 20 hidden channels has only 920 trainable parameters (also factoring in both weights and biases). The GAT models were hence enabled to reach significantly simpler configurations, in the Bayesian optimisations, if necessary for data-scarce learning applications.

While this comparison between the implied expressiveness of the GIN and GAT models is simplistic and focuses purely on the message passing mechanisms, it demonstrates that the allowed upper bound size of the message passing mechanism was significantly more complex for the GINs, compared to the GATs. The GINs used in this study may hence have been more expressive than the GATs and GCNs, yet required larger numbers of training data points, to sufficiently train a larger set of trainable parameters; hence this may explain the observed phenomenon of the GINs marginally outperforming over the majority of assay datasets (which may *ad-hoc* be speculated as having contained sufficient numbers of training data points), while significantly underperforming against GATs over the 2 most data-scarce assays. The smaller maximum number of trainable parameters of the GATs may have conversely proved advantageous for learning over data-scarce environments, with fewer training examples required to build decisive models, utilising the highly efficient self-attention learning mechanism (with a risk of overfitting present, albeit mitigated via the use of 5-fold cross validation).

**3.1.2.2 – Implications of Explored Hyperparameter Ranges for GATs and GINs**

The GATs may have been allowed to reach a comparable implied expressiveness to the GIN models, in terms of number of trainable parameters, had the maximum number of attention heads and hidden channels in Table 1 been substantially increased. Increasing the number of attention heads however would have not necessarily improved model performance by any significant degree, given that each new attention head merely represents a parallel computation of attention weights, intended to garner novel insights into local node neighbourhoods of small molecular graphs; the number of possible useful insights into inter-node relationships is likely to be limited, within the scope of an inherently limited number of attention weights for a given head. Furthermore, inclusion of too many attention heads would have risked over-smoothing the final node features (obtained as an average of all aggregated attention head outputs), as well as potentially a less efficient Bayesian optimisation, through excessive widening of the hyperparameter search space. It may also be similarly argued that increasing the number of hidden channels would have likewise been of inherently limited benefit, given that the hidden channels were used to mine insights from 8 input features per node. It is common practice (albeit a flawed assumption) in deep learning studies to use hidden layers that are of a size *~2n+1* larger than an input layer of size *n* [43]; with a hidden layer size of greater than *2n+1*, furthermore with multiple hidden layers, the expressiveness of the deep neural network to model highly non-linear functions significantly increases. The allowed number of hidden channels, in Table 1, ranges from 20-200, which for 8 input features vastly surpasses the informal *2n+1* guideline for configuring hidden layer sizes (and additionally, the existence of 3-7 GAT layers represents multiple hidden layers) – hence the expressiveness of the GAT models may still be speculated as having been highly sufficient for modelling any patterns and relationships arising from local neighbourhoods of nodes containing 8 physicochemical node features, in molecular graphs of small drug-like molecules.

Given the informal *2n+1* rule for configuring hidden layer sizes, it may be argued that all GNN architectures were enabled to contain *excessive* expressiveness for the modelling tasks at hand – an idea which is supported by the generally close-proximity results of Fig. 5a, with only marginal supremacy of particular GNN architectures demonstrated. A generous range of potential expressiveness was enabled in the hyperparameter search spaces of Table 1, for each GNN, as it was speculated that a sufficient Bayesian optimisation would be capable of converging to an optimal hyperparameter state, utilising an appropriate level of expressiveness, over each dataset. Further examination of this idea requires greater examination of the Bayesian optimisation data, discussed in further sections.

It should finally be noted that the use of an $MLP_{GIN}$ in each GIN layer, containing between 3-5 layers, represented *greater expressiveness within each message passing iteration*, in a way which the utilised GAT architecture was inherently incapable of emulating (with a de-facto depth of only 1 layer used within each message passing iteration). Given the argument of each GNN's hyperparameter search space having likely overall enabled sufficient expressiveness for this study's molecular modelling tasks, further supported by the close proximity of mean peak AUC scores in Fig. 5a, it may hence be postulated that the marginal yet consistent performance differences between the GNNs are indicative of inherent advantages and disadvantages of the underlying message passing algorithms.

The results of Fig. 5, while statistically uncertain, are hence indicative of GATs as an inherently more efficient and effective GNN architecture for data-scarce learning applications in molecular modelling, whereas GINs are capable of more expressive molecular modelling, albeit with a need for considerably more data-abundant environments.

### 3.2 – Correlations of Hyperparameters with Performance

### 3.2.1 – GCN Correlational Bar Chart

Pearson Correlation Coefficients (PCCs) of explored GCN hyperparameters, against mean peak AUC scores, are displayed in Fig. 6. Fig. 6 indicates overall positive correlations for number of GNN layers, number of hidden channels and learning rate, while displaying statistically insignificant towards weakly negative correlations across other hyperparameters.

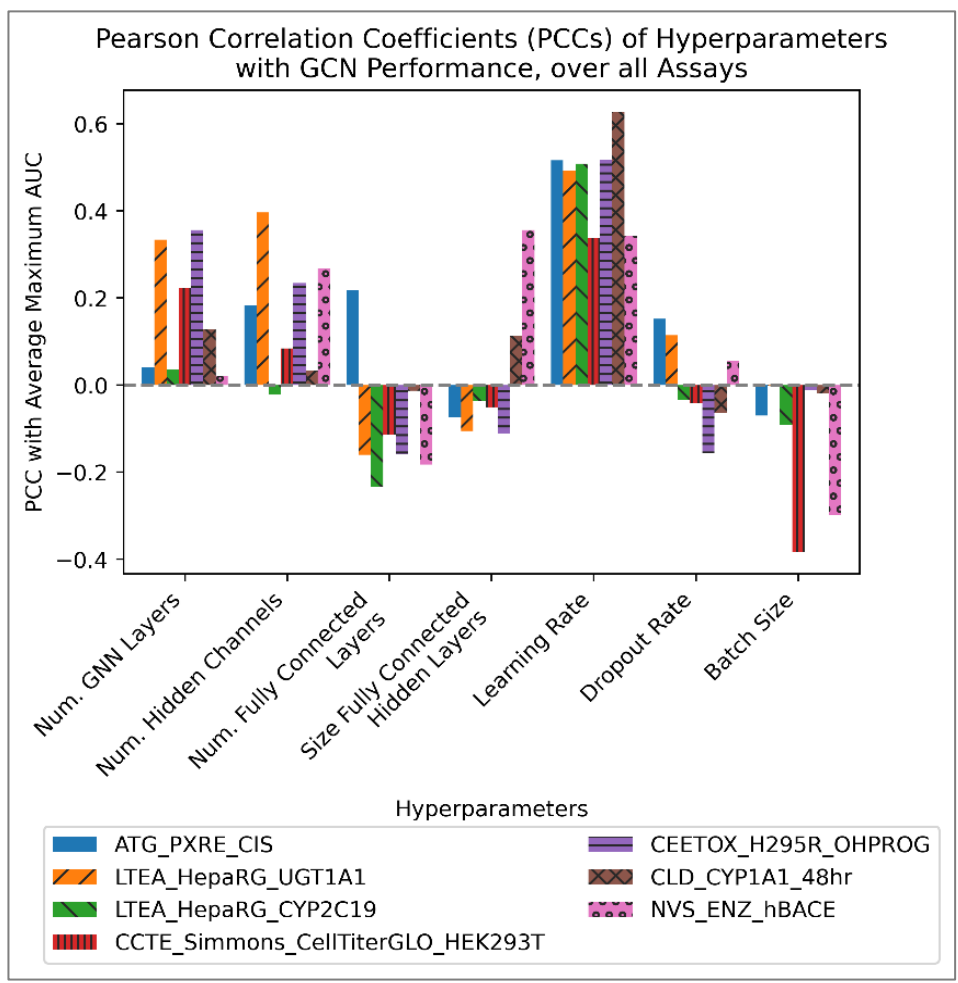

*Figure 6: Bar chart displaying PCCs of explored GCN hyperparameters, against mean peak AUC scores of the model, for the Bayesian optimisations, compared across different toxicological assay datasets (specified in the bar chart key, presented in descending order of size).*

Fig. 6 shows that numerous hyperparameters were positively correlated with mean peak AUC, such as the number of GNN layers, the number of hidden channels and the learning rate. The learning rate was especially strong in its positive correlations, with PCC values of >0.5 across all datasets, implying that vigour of the learning process, in exploring the trainable parameters landscape for test loss minima (and inferred AUC maxima), was more important than precision or the ability of the model to converge into finer optimal regions. The number of GNN layers and number of hidden channels were less strong or consistent in their positive correlations, across different datasets, implying that certain datasets may have been handled better by simpler models, or otherwise hindered by excessively expressive models. There is however no discernible trend in Fig. 6, for weaker PCCs across more data-scarce learning environments, for these two hyperparameters.

Numerous negative correlations are also apparent from Fig. 6, for the number of fully connected layers, size of fully connected layers, dropout rate and batch size. These trends however are notably inconsistent across different datasets considered, hence the ability to draw significant insights is limited.

Fig. 6 should be regarded as merely a superficial initial attempt to gauge the behaviour of the hyperparameter space, as it de-facto relies on the treatment of each hyperparameter as an independent variable that can be uniquely correlated with mean peak AUC. The opposite is in fact true, given that the Bayesian optimisations attempted to vary and optimise all parameters simultaneously, across different iterations, for maximising mean peak AUC; hence the mean peak AUC at a given hyperparameter value is likely to be significantly impacted by other chosen hyperparameter values. At best, Fig. 6 can be regarded as a statistically noisy attempt to garner insights into the hyperparameter space, albeit subject to significant assumptions and uncertainties, even if certain trends in Fig. 6 appear to follow.

### 3.2.2 – GAT Correlational Bar Chart

The same analysis of Fig. 6 is applied to GATs in Fig. 7, revealing notably different trends in correlations of hyperparameters to ROC AUC score. Numerous inconsistent or statistically insignificant PCC trends were uncovered, in addition to more consistent negative correlations over certain hyperparameters.

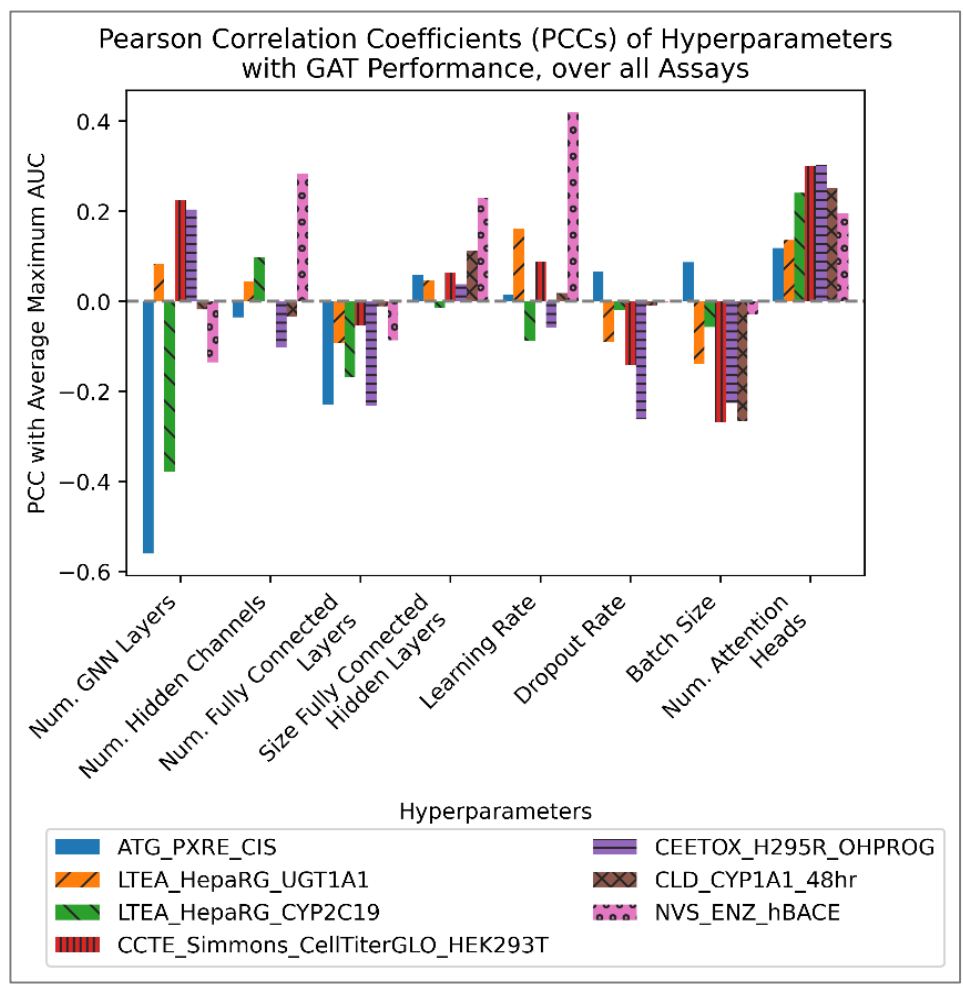

*Figure 7: Bar chart displaying PCCs of explored GAT hyperparameters, against mean peak AUC scores of the model, for the Bayesian optimisations, compared across different toxicological assay datasets (specified in the bar chart key, presented in descending order of size).*

Fig. 7 displays mostly contrary (or otherwise indecisive) trends to those of Fig. 6, for individual hyperparameter correlations with mean peak AUC. This may suggest highly different behaviour of the GAT hyperparameter space, versus the GCN hyperparameter space. However, such a phenomenon is unlikely – both GNN architectures share similarities in their message-passing operations (i.e. the summation of dot products between weights matrices and node feature vectors, as per Equations 1 and 2), with the distinction being the inclusion of self-attention mechanism in GATs; additionally, both share a common architecture framework (Fig. 4).

Furthermore, many of the PCCs of hyperparameters in Fig. 7 are heavily inconsistent, fluctuating between positive and negative values across different datasets. This further affirms the notion that the underlying assumptions of these bar charts are limited in validity; any real underlying correlations between hyperparameters and mean peak AUC are likely obscured by the inherent variability introduced by Bayesian optimisations, which explores a complex and multi-dimensional parameter space.

Despite this, the number of attention heads was consistently positively correlated with mean peak AUC, albeit weakly. This is of intuitive sense, as greater numbers of attention heads enable GATs to compute a broader set of paralleled attention-weighted rpresentations, which can benefit learning from complex graph-structured molecular data. The weak nature of the correlations however may stem from the inherent variability of the Bayesian optimisation and interaction with other relevant hyperparameters, in addition to potential over-smoothing from excessive numbers of attention heads.

Shared trends between Fig. 6 and Fig. 7 include weakly negative PCCs for number of fully connected layers, as well as overall negative (albeit not very consistent) PCCs for batch size.

### 3.2.3 – GIN Correlational Bar Chart

The PCC analysis of ROC AUC score against hyperparameter values was also applied to GINs, in Fig. 8. Fig. 8 reveals further deviation from the trends apparent in Fig. 6 and Fig. 7, albeit with comparably stronger negative correlations apparent (especially for the number of $MLP_{GIN}$ layers).

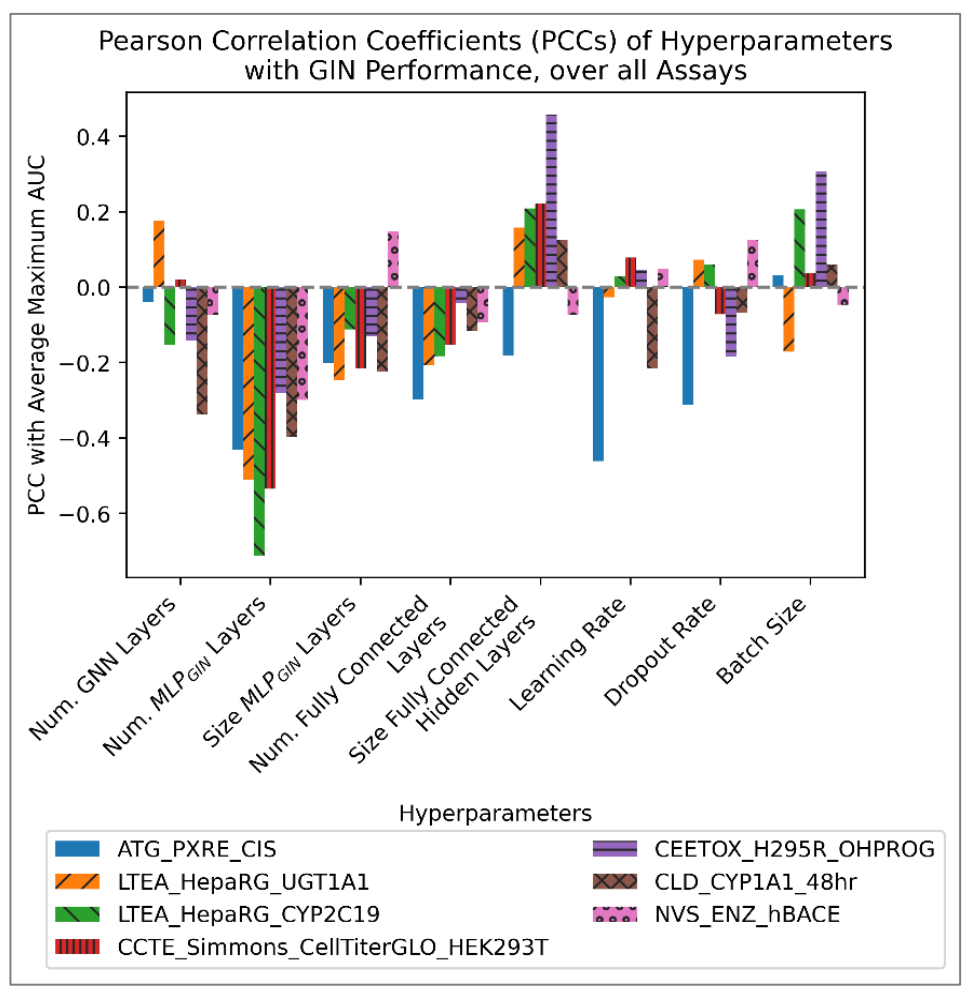

*Figure 8: Bar chart displaying PCCs of explored GIN hyperparameters, against mean peak AUC scores of the model, for the Bayesian optimisations, compared across different toxicological assay datasets (specified in the bar chart key, presented in descending order of size).*

Fig. 8 displays several consistent trends in hyperparameter correlation with mean peak AUC for the GIN models. Most notably, for the number of $MLP_{GIN}$ layers shows strongly negative PCCs, while weak negative PCCs were observed for the number of GNN layers, size of the $MLP_{GIN}$ layers and number of fully connected layers. In contrast, the size of the fully connected hidden layers shows weakly positive PCCs

The strongly negative PCCs for the number of $MLP_{GIN}$ layers aligns with the previous discussion of the high expressiveness of the $MLP_{GIN}$ layers, in Section 3.1.2; inclusion of too many of these layers may have hence been of hinderance to the model's ability to train sufficiently within 500 epochs. A similar argument can also be inferred for the weakly negative PCCs for the size of the $MLP_{GIN}$ layers, as well as for the number of GIN layers, potentially contributing to excessively large model expressiveness, which may worsen the risks of overfitting or lack of efficient converge to sufficient optima.

The weakly negative PCCs for the number of fully connected layers, as well as weakly positive PCCs for their size, initially appear to be contradictory, but may indicate a consistent preference of Bayesian optimisations to achieve expressiveness through wider rather than deeper MLP architectures. This trend is interestingly also vaguely apparent from Fig. 7 for the GAT, albeit inconsistent with the trends of Fig. 6 for the GCN, where both number and size of fully connected layers displayed negative PCCs.

The batch size was of inconsistent yet overall positive PCCs, in Fig. 8, in contrast to the overall negative PCC trends of Fig. 6 and Fig. 7.

### 3.2.4 – Summary of Correlational Bar Charts

The results from Fig. 6, Fig. 7 and Fig. 8 indicate some trends of hyperparameter values displaying positive or negative correlations with mean peak AUC, albeit with high levels of inconsistency between different datasets and GNNs. These inconclusive and statistically high-noise results hence overall affirm the notion of any underlying correlations between hyperparameters and mean peak AUC having been heavily obscured by the chaotic nature of simultaneous hyperparameter variation over Bayesian optimisation iterations, as well as sensitivity of the Bayesian optimisation trajectories (through a high-dimensional hyperparameter space) to initial starting conditions. Any identified trends are likely to either be false as mere coincidences in noisy data, or otherwise true yet indistinguishable from coincidences. The findings from Fig. 6, Fig. 7 and Fig. 8 need not be entirely dismissed, but should be regarded as indicative only, with only superficial and speculative insights enabled.

A more rigorous exploration of the Bayesian optimisations and hyperparameter optimisation landscapes, would hence require simultaneous consideration of all hyperparameter values in conjunction, rather than isolated as purely independent variables of each other (these hyperparameters are independent variables in reality, but cannot be intuitively treated as independent variables in the context of analysing Bayesian optimisation data, where a Bayesian optimisation has varied all hyperparameters simultaneously, to converge towards a maximised mean peak AUC).

### 3.3 – Analysis of Explored Hyperparameter Space

### 3.3.1 – Reduced Dimensionality Hyperparameter Spaces, Between Different GNNs

The Bayesian optimisation searches, over higher-dimensional hyperparameter spaces (10-dimensional, as per Table 1), were compared between the 3 different GNN architectures, over each dataset as a control variable. Given the difficulty of visualising 10-dimensional hyperparameter spaces, this was done by performing Principal Component Analysis (PCA) on the aggregate of the Bayesian optimisation search spaces, for all 3 GNN architectures over each controlled dataset.

Note that, as per Table 1, not all GNN architectures had the same dimensionality of hyperparameter search space (e.g. GCNs did not undergo optimisation for number of attention heads or number or size of $MLP_{GIN}$ layers, as the GCN architecture inherently does not involve such hyperparameters); the explored GCN hyperparameter space had a dimensionality of 7, while the GAT hyperparameter space and GIN hyperparameter space both had dimensionalities of 8, while 10 unique hyperparameter types were explored across all GNN architectures. To directly compare all GNN hyperparameter search spaces, all were first processed into 10-dimensional spaces, via inclusion of all explored hyperparameters and setting values of non-existent hyperparameters for a given GNN to 0. Following this, the 10-dimensional hyperparameter spaces were scaled via the StandardScaler algorithm of the scikit-learn Python library [44], which normalises each column of an input matrix to have a mean of 0 and a unit variance of 1 (this ensures that the PCA algorithm will conserve the true underlying structure of the data, rather than merely differences in scales between columns). PCA was then used to transform the 10-dimensional hyperparameter spaces (aggregated for all GNNs, across a given dataset) into easily visualisable 2-dimensional spaces. Subsequently, comparative 2D scatter plots of Bayesian optimisation search spaces were possible, between all 3 GNNs, per controlled dataset.

For evaluating similarity between optimised hyperparameter states of each GNN in the aggregated hyperparameter space, Euclidean distances were calculated between the most optimal points, in the 2D PCA-transformed hyperparameter space. To facilitate valid comparisons between different calculated Euclidean distances across different 2D PCA-transformed spaces, the Euclidean distances were normalised against the largest distance between any two points in each given 2D PCA-transformed space. This normalised distance, referred to in this study as $d_n$, hence has a theoretical minimum value of 0 and a maximum value of 1. A mathematical formalisation of $d_n$ is provided below:

The set $\mathcal{P}$ of $m$ data points $\boldsymbol{p_i}$, in *M*-dimensional space, may be formalised as:

$$\mathcal{P} = \{\boldsymbol{p_1}, \boldsymbol{p_2}, \ldots, \boldsymbol{p_m}\} \subset \mathbb{R}^M \qquad (7)$$

The Euclidean distance $d_o$ between 2 optima, $\boldsymbol{o_1}$ and $\boldsymbol{o_2}$, may be defined as:

$$d_o = |\boldsymbol{o_1} - \boldsymbol{o_2}| \quad , \qquad \boldsymbol{o_1}, \boldsymbol{o_2} \in \mathcal{P} \qquad (8)$$

The maximum distance $d_{max}$ between any 2 points is:

$$d_{max} = \max_{i,j} |\boldsymbol{p_i} - \boldsymbol{p_j}| \quad , \qquad \boldsymbol{p_i}, \boldsymbol{p_j} \in \mathcal{P} \qquad (9)$$

The normalised distance $d_n$ between the optima, may then be calculated as:

$$d_n = \frac{d_o}{d_{max}} \qquad (10)$$

Scatter plots visualising 2D PCA-transformed hyperparameter spaces, for comparing the Bayesian optimisations of all 3 GNNs across different datasets, are provided below, along with calculated values of $d_n$, as per Equations 7-10:

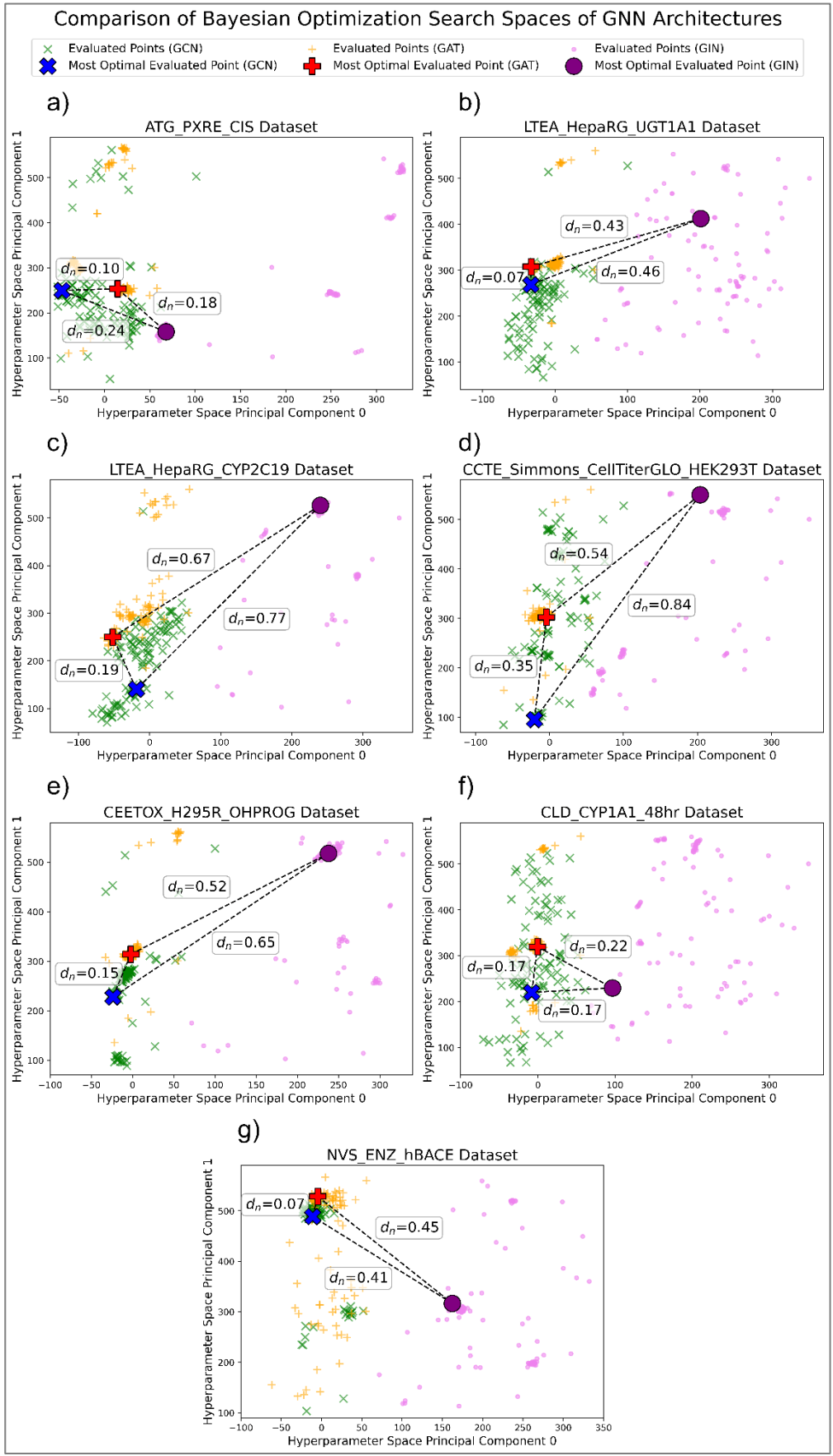

*Figure 9: Scatter plots visualising 2D PCA-transformed hyperparameter spaces, with calculations of $d_n$, for comparing the Bayesian optimisations of all 3 GNNs across different*

*datasets: a) ATG_PXRE_CIS, b) LTEA_HepaRG_UGT1A1, c) LTEA_HepaRG_CYP2C19, d) CCTE_Simmons_CellTiterGLO_HEK293T, e) CEETOX_H295R_OHPROG, f) CLD_CYP1A1_48hr and g) NVS_ENZ_hBACE.*

Fig. 9 displays a consistent trend of larger values of $d_n$ between the optimised GINs and the GCNs, as well as between the GINs and GATs, compared to the smaller $d_n$ values between GCNs and GATs. This indicates that the optimised hyperparameter states of the GINs were measurably less similar than the optimised states of the GCNs and GATs. While this may imply that the GINs reached naturally differing optimised hyperparameter states, across the majority of hyperparameters shared with GCNs and GATs (while also implying that the GCNs and GATs reached naturally closer optimised hyperparameter states), this would assume a fair representation of all hyperparameters in the 2D PCA-transformed spaces.

This assumption may partially hold, given that the PCA algorithm transforms higher dimensional spaces into lower dimensional spaces, in a way that attempts to maximally conserve variance in the original data [45] – the hyperparameter types not shared between all GNN architectures, namely the number of hidden channels, number of attention heads, number of MLP$_{GIN}$ layers and size of MLP$_{GIN}$ layers, would have been of reduced variance, due to presence of uniform 0-values across non-applicable GNN architectures. However, it should be noted that 4 out of 10 of the explored hyperparameters from Table 1, were inapplicable to at least 1 GNN architecture, hence representing 40% of axes of the initial 10-dimensional hyperparameter search spaces. It can hence be assumed that at least some sizable influence of 40% of these axes (which although of impaired variance, would still have contained significant variance across applicable GNNs, due to exploration via the Bayesian optimisation algorithm) would have carried through into the 2D PCA-transformed hyperparameter spaces of Fig. 9. Given that GINs only shared 6 of the explored hyperparameters with GCNs and GATs, while GCNs and GATs shared 7 of the explored hyperparameters with each other, it is reasonable to infer that the greater values of $d_n$ are at least partially due to underlying differences in the types of hyperparameters that were explored. It is however also reasonable to suggest that GINs hold the most inherently different message passing algorithm, compared to GCNs and GATs, and hence would naturally require a naturally more unique set of hyperparameter values (even when controlling for shared hyperparameter types) to optimally navigate the same dataset as GCNs and GATs. A more conclusive investigation of this would require direct comparison of original hyperparameter values, devoid of PCA-transformation.

It should be further noted that use of PCA relies on the assumption that relationships between different variables in the input data (i.e. hyperparameters) are linear in their relationships with each other. This is likely to be a flawed assumption, given that the explored hyperparameters are notoriously linked to AI model performance (in a typically non-linear way) and were simultaneously varied by a non-linear Bayesian optimisation algorithm attempting to maximise model performance. It is however certain that the 10-dimensional hyperparameter optimisation spaces were linearly separable, for partitioning the 3 GNNs, due to the use of 0-values for non-mutual hyperparameter types, which could be used to perfectly distinguish between all 3 GNN architectures in the 10-dimensional aggregate data.

Despite the above discussed implications and limitations of Fig. 9, insights may be considered from Fig. 9a and Fig. 9f, where values of $d_n$ associated with the GINs were significantly smaller than those in all other subplots. While this may imply that the optimised GINs reached closer hyperparameter states to those of the GCNs and GATs, for the associated ATG_PXRE_CIS and CLD_CYP1A1_48hr datasets, it may instead be indicative of the size of the wider hyperparameter spaces explored by the Bayesian optimisation algorithm. As per Equation 10, $d_n$ may be reduced either via reducing $d_o$ or otherwise via increasing $d_{max}$. The ranges of the axes of Fig. 9a and Fig. 9f do not however appear to be substantially different from those of other subplots, nor do the expanses of the associated evaluated points of each constituent Bayesian optimisation appear to be broader – yet the optimal evaluated points do appear to be physically nearer, in the contexts of their wider sets of otherwise ordinarily distributed background evaluated points. It is hence feasible that the GINs did reach closer optimised states to the GCNs and GATs, across the ATG_PXRE_CIS and CLD_CYP1A1_48hr datasets – for reasons which are not clear to the authors of this study, excluding the high likelihood of it being coincidental phenomena of a chaotic system.

Furthermore, throughout the subplots of Fig. 9, it is apparent that certain GNN architectures underwent more adventurous (i.e. wider-spanned) Bayesian optimisation search trajectories than others, likely owing to how rapidly the optimisations were able to converge onto apparent optima in the hyperparameter space. In particular, GATs underwent consistently more spatially limited Bayesian optimisation search trajectories, for reasons which are not clear to the authors of this study. Likewise, the GIN Bayesian optimisation evaluated points appear to have tightly clustered in Fig. 9a and Fig. 9c, as well as to lesser extents in Fig. 9d and Fig. 9e. Overall, the GCN Bayesian optimisations displayed the least tendency to tightly cluster evaluated points, with exceptions in Fig. 9e and Fig. 9g. The reasons for these behaviours, across specific datasets and for specific architectures, are not clear to the authors of this study and are speculated as merely chaotic behaviours by the Bayesian optimisation algorithms, with search trajectories that are highly sensitive to initial conditions and have a tendency to occasionally cluster around sharp optima, in a highly complex and high-dimensional hyperparameter optimisation landscape.

### 3.3.2 – Direct Analysis of 10-Dimensional Hyperparameter Spaces, Between Different GNNs

To address any potential limitations of the PCA-transformations used in Fig. 9, direct analysis of the 10-dimensional hyperparameter optimisation space was carried out. While difficult to directly visualise, normalised Euclidean distances between different models were calculable via direct application of Equations 7-10 (which are compatible for any arbitrary dimensionality of data). In this case, computations of $d_n$ were carried out across the entire aggregate of all 10-dimensional hyperparameter optimisation spaces (aggregated across all GNN architectures and datasets), following min-max normalisation of each column (i.e. total set of explored values for a specific hyperparameter). This enabled direct comparison of all optimised models, regardless of underlying GNN architecture or dataset.

A heat map of this wider comparison, directly over the entire untransformed 10-dimensional hyperparameter optimisation space, is presented below:

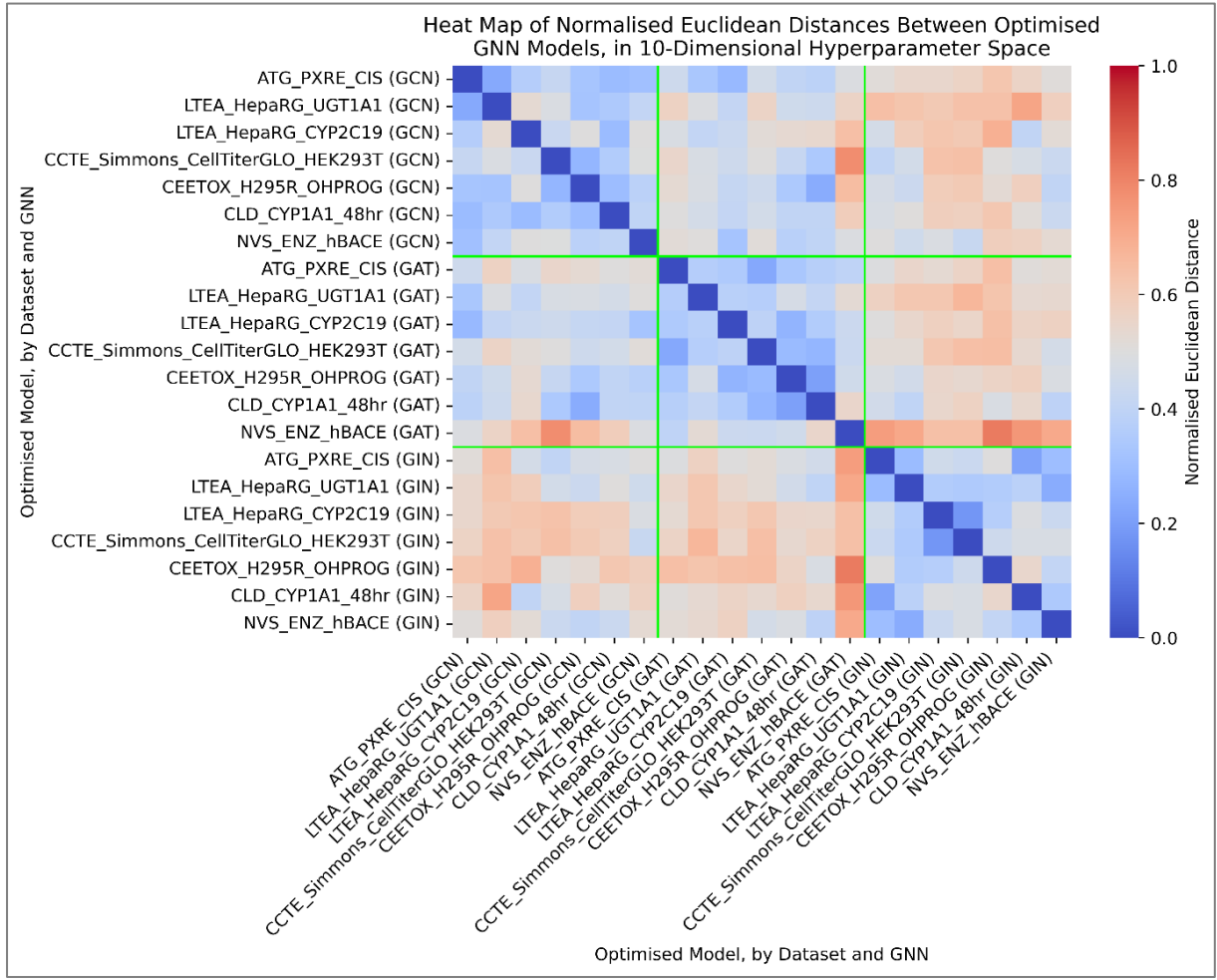

*Figure 10: Heat map of normalised Euclidean distances, over the entire aggregated 10-dimensional hyperparameter space, between optimised models produced in this study. The boundaries between different GNN architectures are marked by green gridlines, to aid in interpretability for the reader.*

Fig. 10 somewhat affirms the trends of Fig. 9 – of GINs being further from GCNs and GATs in the 10-dimensional hyperparameter space. While this may still be due to inherent differences in applicable hyperparameters, the 10-dimensional Euclidean distance analysis, over the raw untransformed 10-dimensional hyperparameter optimisation space, eliminates any of the speculated limitations of using PCA-transformations. The differences in applicable hyperparameters would have contributed to larger distances, however the extent of its impact on Fig. 10 may be estimated; Euclidean distance between 2 points, for any *M*-dimensional space, may be evaluated via:

Let:

$$x = (x_1, x_2, \dots, x_M), \quad y = (y_1, y_2, \dots, y_M) \quad (11)$$

Euclidean distance between $x$ and $y$ is:

$$d(x, y) = \sqrt{\sum_{i=1}^{M}(x_i - y_i)^2} \quad (12)$$

Since certain hyperparameters are not shared between the GIN and other GNNs, and hence entail artificially set zero-values, against larger non-zero values, we can estimate each $(x_i - y_i)^2$ for these affected axes to be ~1 (assuming that min-max scaling set the larger non-zero values to close to 1). For comparing the GIN to the GCN, 3 axes may be approximated this way (number of hidden channels, number of $MLP_{GIN}$ layers and size of $MLP_{GIN}$ layers, while both the GCN and GIN share identical zero-values for the number of attention heads). For comparing the GIN to the GAT, 4 axes may be approximated this way (number of hidden channels, number of $MLP_{GIN}$ layers, size of $MLP_{GIN}$ layers and number of attention heads).

Percentage impact on the normalised Euclidean distances may be estimated, by considering minimum and maximum distances allowed, under these constraints. For minimum constraints, assuming that the GINs shared all other exact hyperparameter values with the GCNs and GATs (hence distances of 0 along these axes), minimum Euclidean distance via Equation 12 is $\sqrt{3}$ between the GINs and GCNs, while $\sqrt{4} = 2$ between the GINs and GATs. For maximum constraints (where all $(x_i - y_i)^2 = 1$), Euclidean distances may be at a maximum of $\sqrt{10}$. Hence the approximate minimum contribution of the non-shared hyperparameter values, as a proportion of the maximum possible Euclidean distance, is $\frac{\sqrt{3}}{\sqrt{10}}$ between the GINs and GCNs, while $\frac{2}{\sqrt{10}}$ between the GINs and GATs, representing respective percentages of 54.8% (3 s.f.) and 63.2% (3.s.f). Between the GCNs and GATs, only 1 hyperparameter (number of attention heads) served as a comparable axis contributing inherent distance, contributing to an estimated proportion of $\frac{1}{\sqrt{10}}$ (i.e. 31.6%; 3 s.f.) of their Euclidean distance. The impact of the non-shared hyperparameter values, on causing larger Euclidean distances for GINs in the 10-dimensional hyperparameter space, may hence be deemed as significant and a leading cause of the results displayed in Fig. 9 and Fig. 10.

It further follows that the approximate impact of non-shared hyperparameters was greater on calculated Euclidean distances between the optimised GINs and GATs, than between the GINs and GCNs, as per Fig. 10; the regions of the heat map which show Euclidean distances between GINs and GCNs, have greater numbers of blue (i.e. nearer) or neutral squares, compared to the regions showing Euclidean distances between GINs and GATs. Interestingly, however, the opposite trend is generally indicated in Fig. 9, although the direct analysis of the 10-dimensional hyperparameter space of Fig. 10, devoid of PCA-transformations, may be deemed as more strongly conclusive.

While excluding non-shared hyperparameters from these analyses may be considered, to address the extensively outlined limitations, such an approach would risk falsely treating the shared hyperparameters as independent of the excluded ones (e.g regarding the number of GIN layers as independent from the number of $MLP_{GIN}$ layers, when in fact both are non-linearly related to peak mean AUC score and were simultaneously varied by a non-linear Bayesian optimisation algorithm, to maximise this). A more robust comparison of different optimised GNN architectures may hence be best approached via direct comparisons of individual hyperparameter values.

A separate interesting observation from Fig. 10 is the prominent distance of the NVS_ENZ_hBACE (GAT) model from the GCN models (and to a lesser extent, other GAT models). There is however no apparent reason for this behaviour, and it may simply be an anomalous outcome from a chaotic optimisation process, over a highly complex and high-dimensional hyperparameter optimisation landscape.

### 3.4 – Direct Comparison of Optimised Model Hyperparameter States

Radar charts, directly comparing optimised hyperparameter configurations, are displayed below:

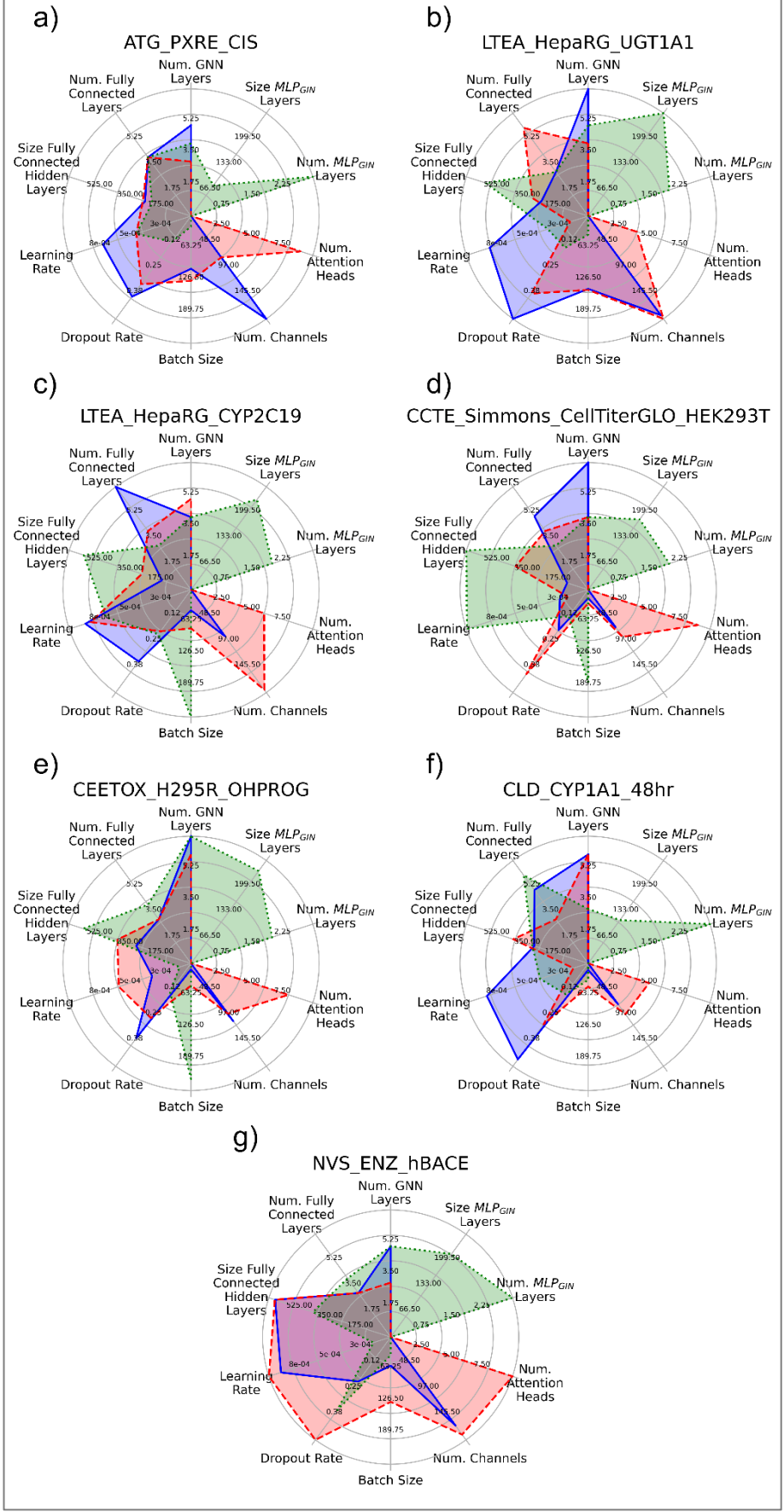

*Figure 11: Radar charts of optimised hyperparameter configurations, for GCNs, GATs and GINs, over all datasets: datasets: a) ATG_PXRE_CIS, b) LTEA_HepaRG_UGT1A1, c) LTEA_HepaRG_CYP2C19, d) CCTE_Simmons_CellTiterGLO_HEK293T, e) CEETOX_H295R_OHPROG, f) CLD_CYP1A1_48hr and g) NVS_ENZ_hBACE.*

Fig. 11 reveals significant deviation between different optimised hyperparameter configurations, both when comparing for the same GNN architectures across different datasets, as well as when comparing between GCNs, GATs and GINs over the same given dataset. While certain trends may be apparent, such as comparatively larger numbers of GNN layers for the optimised GCN models, as well as comparatively higher dropout rates for both the GCN and GAT models, potential explanations of such findings are inherently *ad-hoc* and subject to considerable inconsistencies (e.g. Fig. 11c challenges the notion of GCN numbers of GNN layers being higher than for the other 2 optimised architectures, while Fig. 11d and Fig. 11g contradict the trend of higher dropout rates for the GCNs, whereas Fig. 11c and Fig. 11e likewise contradict the trend of higher dropout rates for the GATs). Other apparent similarities in radar plot shapes in Fig. 11 are due to dips (to 0) for inapplicable hyperparameters (number of hidden channels, number of attention heads, number of $MLP_{GIN}$ layers and/or size of $MLP_{GIN}$ layers, as per Table 1).

Fig. 11 does however appear to demonstrate a limited number of reliable trends, such as sizable numbers of attention heads for the optimised GATs (consistently between 4 and the maximum bound of 9 – often at values in between, while evasive of the allowed minimum of 2). This is indicative of sufficient numbers of attention heads (i.e. sufficient strength of the GAT message passing mechanism) having been of importance to the optimised configurations, which is an expected finding.

Furthermore, the numbers of $MLP_{GIN}$ layers for the optimised GINs were at 2 layers, with relative consistency, across Fig. 11b, Fig. 11c, Fig. 11d and Fig. 11e, with the remaining 3 plots at 3 $MLP_{GIN}$ layers – despite up to 5 layers having been allowed in Table 1. This relatively consistent number of 2-3 $MLP_{GIN}$ layers indicates that the $MLP_{GIN}$ message passing mechanisms of the GINs benefitted from a level of simplicity, below the comparatively large expressiveness allowed by the search boundaries of Table 1. Closer examination also reveals what appears to be a negative correlated relationship between the number and size of the $MLP_{GIN}$ layers, especially across Fig. 11a, Fig. 11b, Fig. 11c, Fig. 11e and Fig. 11f; larger sizes of $MLP_{GIN}$ layers appear to have been configured when using just 2 $MLP_{GIN}$ layers, while smaller sizes were configured when using 3 $MLP_{GIN}$ layers. This apparent trend implies that both the number and size of the $MLP_{GIN}$ layers contributed similarly to reaching the necessary expressiveness, but that the exact allocation of numbers of layers versus sizes of layers was somewhat arbitrary (subject to the Bayesian optimisation trajectory) and that a maximum desirable GIN layer expressiveness was present and converged upon via the Bayesian optimisation algorithms.

Despite trends relating to the message passing mechanisms being discernible for the GATs and GINs, relatively less decisive trends were present for the number of hidden channels for the GCNs, as well as the hidden channels for the GATs, although in both cases the numbers of hidden channels were notably larger than the allowed minimum of 20. A more decisive trend is that the optimised GCNs and GATs displayed closely comparable numbers of hidden channels across Fig. 11b, Fig. 11d, Fig. 11e, Fig. 11f and Fig. 11g. Given the relative similarity between the message passing of both GCNs and GATs, of which both entail use of hidden channels, while also subject to considerable differences, this finding hence supports the notion that approximate global optima (or otherwise local optima with comparable implications for message passing) were converged upon by the Bayesian optimisations, despite other findings of this study having suggested statistically high-noise and chaotic Bayesian optimisation trajectories.

Fig. 11 demonstrates only a limited number of statistically significant trends between different optimised hyperparameter configurations over different GNNs and datasets. This is however still a significant finding, as it supports the notion that the Bayesian optimisations were highly chaotic and sensitive to initial conditions (i.e. randomly initialised starting trainable parameter values, as well as specific GNNs, datasets and fold-configurations). It should however be noted that some shared characteristics emerged between local optima, across models with similarities (e.g. similar numbers of hidden channels, between the GCNs and GATs). Despite some limited similarities between different optimised GNN architectures over different datasets, it may overall be inferred that a specific GNN model, trained on a specific dataset, has a highly unique Bayesian optimisation trajectory and likely converges to a unique local optimum (albeit with some shared traits with other local optima). It may be reasoned that the hyperparameter optimisation landscapes are inherently unique for each GNN-dataset pairing, with differing layouts of global optima. It is unlikely that the Bayesian optimisations converged to global optima, given the complexity of the optimisation landscapes, as well as the underlying constraints (e.g. search space limitations of Table 1).

The Bayesian optimisations may have been improved, in their abilities to converge towards global optima, by allowing for larger numbers of samples (e.g. 300 evaluated points, instead of 100), as well as widened search spaces for different hyperparameter values and even larger numbers of considered hyperparameters. However, doing so would have significantly increased computational cost, for potentially only marginal improvements in peak mean AUC scores of optimised models. Hyperparameter optimisation can pragmatically be regarded as an imperfect yet necessary step for optimising models, especially for studies such as this one, which attempt to draw fair comparisons between different models. Although the Bayesian optimisations are always improvable, via greater allocation of computational resources, the optimisations of this study are nonetheless deemed as sufficient and fairly controlled, for the purposes of enabling fair comparisons between different GNN architectures over different toxicological data environments. The highly differing optimal hyperparameter configurations found, presented in Fig. 11, further affirms the benefit of having carried out separate Bayesian optimisations over each dataset, for the purposes of navigating potentially unique hyperparameter optimisation landscapes with unique optima, for enabling fair comparisons between the considered GNN architectures.

## 4 – Conclusions

Through our extensive results and analyses, it may be concluded that the original hypothesis has been largely affirmed; different GNN architectures displayed differing levels of performance, over the varied toxicological assay data environments, when subjected to controlled conditions. The hypothesis that GATs and GINs would outperform GCNs, as comparably more advanced GNN architectures, has been strongly confirmed, consistent across all findings. Furthermore, it was also affirmed that different GNN architectures would display unique strengths and weaknesses over differing data-abundance environments; GATs were found to perform more optimally over data-scarce environments, whereas GINs were found to consistently perform more optimally over data-abundant environments.

The extent of GIN supremacy over data-abundant environments was consistent, although not significantly above GAT performance, compared to the extent of GAT supremacy over GIN performance for data-scarce environments. GATs are hence concluded as a highly effective and versatile algorithm over all data environments, for molecular modelling tasks; highly advantageous for data-scarce learning applications, while also sufficient in performance over data-abundant environments (with a significantly lower computational cost than GINs). GINs however may prove advantageous, when applied to tasks of abundant training data and computational resources and where high importance of even marginal improvements in performance is present.

Furthermore, the hypothesis was affirmed regarding unique optimised hyperparameter configurations, over unique hyperparameter optimisation landscapes of high complexity. The findings of this study indicated that the Bayesian optimisation trajectories were highly chaotic, sensitive to initial conditions and struggled to navigate local optima, in highly complex and unique optimisation landscapes, in the task of converging towards a global optimum. The Bayesian optimisations and subsequent analyses, while imperfect, were nonetheless regarded as sufficient for enabling fair comparisons and scientifically valid conclusions to be drawn. GINs were found to reach more distant optima in hyperparameter space, compared to GCNs and GATs. This naturally follows from their inherent theoretical differences from GCNs and GATs, in terms of message passing algorithm and associated hyperparameters, which our extensive analyses further affirmed.

These findings offer generalisable comparative insights into the behaviour of different GNN architectures across varied data regimes. Although no external validation set was used, as per Section 2.5, the use of rigorous 5-fold cross-validation, coupled with the observation of consistent trends across multiple datasets and fold configurations, supports the notion that the risks and impacts of potential overfitting were successfully mitigated, while also supporting the reliability of the conclusions drawn.

Future improvements to this study could entail more extensive Bayesian optimisations, particularly involving larger numbers of iterations, considered hyperparameters and hyperparameter value search ranges. Through incorporating greater numbers of GNN hyperparameters and wider search ranges, potentially greater model complexities may be reached. Inclusion of external validation data may also be considered, resulting in more computationally expensive nested cross-validation instances, albeit improving applicability of models obtained and computing performance metrics which may be used as reliable benchmarks in further studies.

Additional GNN architectures, such as GraphSAGE, GTVs and GTNs, may also be included in future studies, to enable broader-reaching comparisons over the geometric deep learning architecture space. Additionally, the scope of the study may be expanded, to compare predictive models over regression-based tasks, as well as the classification-based tasks considered in this study. Widening the scope of the study to other cheminformatics-based domains, outside of computational toxicology, may also be considered – e.g. predicting drug efficacy, physical properties in materials science, modelling chemical reactions etc. The wider topic of comparing different geometric deep learning architectures, for cheminformatics-based tasks, may also be expanded beyond predictive models of bioactivity or material properties – e.g. performance in informing generative models, as well as modelling of chemical reactions.

## List of Abbreviations

Artificial Intelligence (AI); Graph Neural Network (GNN); Graph Convolutional Network (GCN); Graph Attention Network (GAT); Graph Isomorphism Network (GIN); Area Under the Curve (AUC); Quantitative Structure-Activity Relationship (QSAR); Convolutional Neural Network (CNN); Rectified Linear Unit (ReLU); Natural Language Processing (NLP); Exponential Linear Unit (ELU); 1-Weisfeiler-Lehman (1-WL); Multi-Layer Perceptron (MLP); Graph Sample and Aggregate (GraphSAGE); Graph Transformer Network (GTN); Graph Total Variation (GTV); Graph Autoencoder (GAE); Message Passing Neural Network (MPNN); SMILES (Simplified Molecular Input Line Entry System); PyTorch Geometric (PyG); Receiver Operating Characteristic - Area Under the Curve (ROC AUC); Pearson Correlation Coefficient (PCC); Principal Component Analysis (PCA).

## Declarations

### Availability of Data and Materials

All software developed, as well as data and other materials, are available in the following GitHub folder (constituent of the wider TOX-AI GitHub repository): https://github.com/alexanderkalian/TOX-AI/tree/main/comparing_gnns

### Competing Interests

The authors of this study declare no competing interests.

### Funding

This work was supported by grants from the Biotechnology and Biological Sciences Research Council [grant number BB/T008709/1] and the Food Standards Agency [Agency Project FS900120].

### Authors' Contributions

ADK conceptualised the study. CH, MG, OJO, EB, CP and JLCMD supervised the study. ADK carried out computational experiments for the study. ADK produced analyses of the results. ADK, LO, JL and MG contributed to writing the manuscript. ADK produced diagrams for the manuscript. ADK, LO, JL, OJO, MG and CH reviewed and edited the manuscript.

### Acknowledgements

The content of this article does not necessarily reflect the views of the Food Standards Agency.

The views expressed in this article do not reflect the views of the European Food Safety Authority (EFSA) and/or are a reflection of the views of the authors only.

This paper aims to contribute to the international network on Advancing the Pace of Chemical Risk Assessment (APCRA), to contribute to the use of New Approach Methodologies (NAMs) in chemical risk assessment and ultimately reduce animal testing.